\documentclass[reprint,preprintnumbers,nofootinbib,amsmath,amssymb,aps,prd]{revtex4-2}

\usepackage{amsmath,amssymb}
\usepackage{graphicx}
\usepackage[caption=false]{subfig}
\usepackage{hyperref}
\graphicspath{{img/}}

\newcommand{\ved}[2]{
\begin{array}{c}
 #1\\
#2
\end{array}
}

\renewcommand{\vec}[1]{\mathbf{#1}}

\newcommand{\abs}[1]{\left\lvert{#1}\right\rvert}
\newcommand{\absl}[1]{\lvert{#1}\rvert}
\newcommand{\avg}[1]{\left\langle{#1}\right\rangle}
\newcommand{\diff}{\mathrm{d}}

\newcommand{\me}{\mathrm{e}}
\newcommand{\Laplace}{\mathop{}\!\mathbin\bigtriangleup}

\begin{document}
\numberwithin{equation}{section}

\title{Probing Dark Matter Clumps, Strings and Domain Walls \\ with Gravitational Wave Detectors}

\author{Joerg Jaeckel}
\affiliation{Institut f\"ur theoretische Physik, Universit\"at Heidelberg, Philosophenweg 16, 69120 Heidelberg, Germany}

\author{Sebastian Schenk}
\affiliation{Institute for Particle Physics Phenomenology, Department of Physics, Durham University, DH1 3LE, United Kingdom}

\author{Michael Spannowsky}
\affiliation{Institute for Particle Physics Phenomenology, Department of Physics, Durham University, DH1 3LE, United Kingdom}

\preprint{IPPP/20/9}

\begin{abstract}
Gravitational wave astronomy has recently emerged as a new way to study our Universe.
In this work, we survey the potential of gravitational wave interferometers to detect macroscopic astrophysical objects comprising the dark matter.
Starting from the well-known case of clumps we expand to cosmic strings and domain walls.
We also consider the sensitivity to measure the dark matter power spectrum on small scales.
Our analysis is based on the fact that these objects, when traversing the vicinity of the detector, will exert a gravitational pull on each node of the interferometer, in turn leading to a differential acceleration and corresponding Doppler signal, that can be measured.
As a prototypical example of a gravitational wave interferometer, we consider signals induced at LISA.
We further extrapolate our results to gravitational wave experiments sensitive in other frequency bands, including ground-based interferometers, such as LIGO, and pulsar timing arrays, e.g.~ones based on the Square Kilometer Array.
Assuming moderate sensitivity improvements beyond the current designs, clumps, strings and domain walls may be within reach of these experiments.
\end{abstract}

\maketitle

\section{Introduction}
\label{sec:introduction}

LIGO's measurement of gravitational waves emitted from black hole binary mergers~\cite{Abramovici:1992ah,Abbott:2016blz} has opened novel ways to study the properties of astrophysical objects via their gravitational interactions.
After decades-lasting developments and technical improvements on interferometrical methods, an unprecedented sensitivity in the measurement of gravitational interactions has been achieved.
Future ground- and space-based interferometers, such as the Cosmic Explorer~\cite{Reitze:2019iox}, the Einstein Telescope~\cite{Punturo:2010zz} and LISA~\cite{Cutler:1997ta,Berti:2005ys}, are expected to further improve on this sensitivity and to complement it in different frequency ranges. In addition, this may also be supplemented by instruments such as AEDGE~\cite{AEDGE:2019nxb}, AION~\cite{Badurina:2019hst}, BBO~\cite{Corbin:2005ny} or DECIGO~\cite{Kawamura:2006up}.
To achieve sensitivity to lower gravitational wave frequencies and smaller strains, longer interferometer arms and less background noise are needed.
In the former, pulsar timing arrays, such as NANOGrav~\cite{Jenet:2009hk}, PPTA~\cite{Hobbs:2009yy} or EPTA~\cite{Ferdman:2010xq}, currently provide additional tests of gravitational physics.
In the future, pulsar timing arrays might even use the Square Kilometer Array (SKA)~\cite{Smits:2008cf} to further improve on sensitivity.

In this work, we want to explore the potential of gravitational wave interferometers to probe concentrated structures of dark matter via their gravitational interaction with the apparatus.
This provides an additional science goal for these instruments, but also requires different analysis strategies.
While previous studies~\cite{Adams:2004pk,Seto:2004zu,Siegel:2007fz,Seto:2007kj,Berezinsky:2010kq,Baghram:2011is,Kashiyama:2012qz,Berezinsky:2014wya,Clark:2015sha,Hall:2016usm,Kawasaki:2018xak,Schutz:2016khr,Kashiyama:2018gsh,Dror:2019twh,Grote:2019uvn} have focused mainly on clumps and primordial black holes, we consider also topological defects, such as cosmic strings and domain walls.
The opportunity to search for the latter two has already been noted in~\cite{Hall:2016usm,Dokuchaev:2017psd}\footnote{In~\cite{Dokuchaev:2017psd} the main focus is on axion streams arising from the destruction of axion miniclusters, which are, however, rather similar to strings.}.
Here, we aim for a more detailed analysis of the experimental sensitivity and also consider the possibility that the topological defects feature a more general equation of state, e.g.~due to a non-trivial behavior of the string/domain wall network.
We also take a brief look beyond localized structures and consider the possibility to measure the local dark matter density fluctuation power spectrum (see also~\cite{Baghram:2011is}).
The sensitivity to the latter is, however, somewhat limited.

Our analysis is based on the simplest effect (originally discussed in~\cite{Seto:2004zu}), namely, that any massive object or, more precisely, any localized energy density will locally perturb the gravitational field in the vicinity of the interferometer, thereby exerting a different gravitational pull on each node of the detector.
Due to the differential gravitational acceleration, this perturbation will, in turn, lead to a measurable Doppler shift in the apparatus.
For instance, a similar analysis of such acceleration burst signals at space-based interferometers has been discussed for nearby asteroids and similar objects~\cite{Vinet:2006fj,Purdue:2007zz,Tricarico:2009zz}.
In addition to this, a signal could also arise due to the (changing) gravitational potential when a structure exists within the line of site connecting the different nodes of the gravitational wave detector~\cite{Siegel:2007fz,Baghram:2011is,Clark:2015sha,Schutz:2016khr,Dror:2019twh}.
It would be interesting to study this Shapiro effect~\cite{Shapiro:1964uw} also for cosmic strings and domain walls.
We leave this for future work.

In principle, this type of dark matter search is very general as it solely relies on measuring purely gravitational interactions\footnote{Signals of dark matter structures in gravitational wave interferometers can also arise from non-gravitational interactions, see, e.g.,~\cite{Stadnik:2014tta,Stadnik:2015xbn,Graham:2015ifn,Jaeckel:2016jlh,Hall:2016usm,Pierce:2018xmy,Grabowska:2018lnd,McNally:2019lcg}.}.
Of course, the downside is that it requires the dark matter to be strongly concentrated in highly localized structures which, in turn, re-introduces a model dependence.
While our study is purely phenomenological and does not rely on the origin of the structures in question, let us nevertheless mention a few possibilities allowing for such scenarios.

Localized clumps in the dark matter structure can appear in a variety of models and situations.
Perhaps the most obvious scenario for clumps are primordial black holes~\cite{Carr:1974nx,Meszaros:1974tb,Carr:1975qj,Ivanov:1994pa,GarciaBellido:1996qt} (in the context of gravitational wave astronomy, see also more recently, e.g.,~\cite{Bird:2016dcv,Garcia-Bellido:2017fdg,Sasaki:2018dmp,Barack:2018yly,Cai:2018dig,Bartolo:2018evs} and~\cite{Carr:2016drx} for a recent review).
That said, already standard cold dark matter may feature clumpy structures (see, e.g.,~\cite{Baghram:2011is,Berezinsky:2014wya}), although these may likely be beyond the reach of near future gravitational wave detectors~\cite{Baghram:2011is}.
In addition, localized clumps can also be produced if dark matter features strong self-interactions~\cite{Kaplan:2009de,Wise:2014ola,Hardy:2014mqa,Gresham:2017zqi}.
Another important potential source are initial conditions featuring large inhomogeneities, a prominent example of which are axion miniclusters~\cite{Hogan:1988mp,Kolb:1993zz,Kolb:1993hw,Zurek:2006sy} (see also~\cite{Fairbairn:2017dmf,Fairbairn:2017sil,Enander:2017ogx,Vaquero:2018tib,Eggemeier:2019khm} for some more recent work).
More recently, such large fluctuations have also been discussed in the context of inflationary production mechanisms~\cite{Graham:2015rva,Alonso-Alvarez:2018tus} or as a consequence of a fragmentation of homogeneous fields due to their (self-)interactions~\cite{Berges:2019dgr,Fonseca:2019ypl}.
Furthermore, macroscopic clumps could arise as solitonic objects such as, e.g., $Q$-balls~\cite{Coleman:1985ki,Lee:1988ag,Kusenko:2001vu,Kusenko:1997si}.

In addition to localized clumps, dark matter structures might also come in form of topological defects, such as cosmic strings or domain walls.
These could have formed in the early Universe~\cite{Kibble:1980mv,Vilenkin:1984ib,Press:1989yh}.
In a cosmological context, topological defects, in particular dynamical networks of cosmic strings or domain walls (see, e.g.,~\cite{Vilenkin:1984rt,Spergel:1996ai,McGraw:1996py,McGraw:1997nx,Bucher:1998mh,Battye:1999eq,Friedland:2002qs}), contribute to the total energy budget of the Universe.
However, while their equation of state is usually negative, there is still uncertainty in the behaviour of a network of these objects (see, e.g.,~\cite{Avelino:2008ve,Sousa:2009is}).
Therefore they may account for dark energy but also for dark matter~\cite{Bucher:1998mh,Battye:1999eq,Friedland:2002qs}.
This has led to a variety of ideas to investigate their experimental signatures as dark matter candidates~\cite{Pospelov:2012mt,Derevianko:2013oaa,Motohashi:2013kka,Stadnik:2014tta,Stadnik:2014ala,Stadnik:2014cea,Jaeckel:2016jlh,Roberts:2017hla,Grote:2019uvn,Masia-Roig:2019hsy,McNally:2019lcg}.
We will follow this more phenomenological spirit and be agnostic about the equation of state of these networks by treating it as a free parameter.
This allows us to investigate their experimental imprint in gravitational wave detectors independent of the dynamics of the network.

Our discussion is structured as follows.
Section~\ref{sec:dmclumps} first reviews the analysis of acceleration burst signals at the LISA gravitational wave interferometer.
As an initial test case, we then apply these techniques to obtain the signal power spectrum associated to localized clumps of dark matter passing by the detector.
In Section~\ref{sec:topologicalobjects}, we extend our analysis to the case of cosmic strings and domain walls.
Next, Section~\ref{sec:dmfluctuations} gives an overview of how the same technique might also be used to measure stochastic fluctuations of the local dark matter density with LISA.
In Section~\ref{sec:GWInterferometers}, we extrapolate our results to other gravitational wave experiments, i.e. ground-based interferometers and pulsar timing arrays, that are sensitive to different frequency bands.
As particular examples, we examine LIGO and a future PTA using the SKA.
Finally, we summarize our results and conclude in Section~\ref{sec:conclusions}.

\section{Localized Clumps of Dark Matter}
\label{sec:dmclumps}

Localized dark matter clumps can cause a signal in a gravitational wave interferometer as shown by a number of previous studies~\cite{Adams:2004pk,Seto:2004zu,Siegel:2007fz,Seto:2007kj,Berezinsky:2010kq,Baghram:2011is,Kashiyama:2012qz,Berezinsky:2014wya,Clark:2015sha,Hall:2016usm,Kawasaki:2018xak,Schutz:2016khr,Kashiyama:2018gsh,Dror:2019twh,Grote:2019uvn}.
In this section, we will review how such a signal is generated.
We focus on the effect where the clump exerts a stronger gravitational acceleration on one of the interferometer nodes and derive the associated signal power spectrum for the LISA detector.
This discussion will then serve as a basis and test bed for our investigation of strings and domain walls in the next section.
For a general introduction to the physics and measurement techniques of gravitational wave interferometers see, e.g.,~\cite{Saulson:1995zi,Maggiore:1900zz}.

Although the general strategy in principle applies to any gravitational wave detector, in the present work, we will consider LISA as a prototypical example to obtain the experimental signature of gravitational perturbations caused by the above macroscopic astrophysical objects in a gravitational wave interferometer.
We will later (rather naively) use the same technique to estimate the sensitivity in other gravitational wave experiments.
Here, loosely speaking, we focus on an experimental setup with three distinct nodes arranged in an equilateral triangle (see Appendix~\ref{app:geometry} for some more details).
Along the three interferometer arms with a length of about 2.5 million kilometers, the satellites exchange laser beams.
Consequently, a differential acceleration due to a gravitational perturbation caused by macroscopic objects in the vicinity of the interferometer will then lead to a measurable Doppler signal.

To begin with, let us first consider the case of a single dark matter clump passing by one of the LISA satellites.
This is similar to the case of the detection of asteroids treated in~\cite{Vinet:2006fj}, which we will follow closely.
As a simple coordinate frame, we choose the dark matter clump to be in a straight uniform motion with velocity $V$ parallel to the $y$-axis.
The trajectory of the clump is, in addition, confined to the $xy$-plane.
The closest distance between the clump and the satellite located at the origin, i.e.~the impact parameter, is denoted by $D$.
For a schematic illustration of this reference frame we refer the reader to Appendix~\ref{app:GeomSpheres}.
The differential acceleration of the interferometer node is then given by~\cite{Vinet:2006fj},
\begin{equation}
	\vec{g}(t) = \frac{GM}{D^2} \frac{1}{\left( 1 + \left(\frac{Vt}{D}\right)^2\right)^{3/2}} \begin{pmatrix} 1 \\ \frac{Vt}{D} \\ 0 \end{pmatrix} \, ,
\label{eq:DMacceleration}
\end{equation}
where $G$ denotes the gravitational constant and $M$ is the mass of the dark matter clump.
A schematic form of this acceleration burst was first considered in~\cite{Seto:2004zu} with regard to the detection of primordial black holes with space-based interferometers.
As~\cite{Vinet:2006fj}, in our analysis we will instead consider the velocity shift, $\vec{v}(t)$, i.e.~the integrated gravitational acceleration.
However, we will use the more appropriate signal response functions defined in~\cite{Armstrong:1999,Dhurandhar:2002zcl} that take into account a time retardation of the signals from different nodes (see also discussion below).

In principle, the gravitational field of the dark matter clump passing by the detector will exert a gravitational pull on each interferometer node separately.
Due to the differential velocity shifts relative to each other, the laser beams exchanged between the satellites will be influenced by Doppler shifts.
That is, we expect a time-dependent response of the detector to these velocity perturbations.
This signal is parametrized by a so-called response function $X(t)$.
The signal power spectrum, that we are interested in, is given by the absolute square of the Fourier transform of the detector response\footnote{Technically, the power spectrum of a given signal $X(t)$ is defined by the Fourier transform of the so-called auto-correlation function, $P (\omega) = \mathcal{F} \left[ \left( X \star X \right) (\tau) \right] = \mathcal{F} \left[ \int_{\mathbb{R}} \diff t \,  X^{\ast}(t) X(t+\tau) \right]$. The latter is, in fact, equivalent to $\absl{\tilde{X}(\omega)}^2$, i.e.~to the definition given in the main text. Throughout this work, we will denote the Fourier transform of a function $f(t)$ by $\tilde{f}(\omega)$.},
\begin{equation}
	P (\omega) = \abs{\tilde{X}(\omega)}^2 \, .
\label{eq:PowerSpectrum}
\end{equation}
In the case of LISA, this response function typically is a linear combination of the velocity perturbations of all three interferometer nodes.
Its exact form depends on which undesirable noise sources are tried to be removed from the signal spectrum.
Therefore, the detector response function is not unique.
For concreteness, throughout this work, we will use the so-called \textit{Michelson} response function $X(t)$ for the readout of a signal at a single detector node~\cite{Armstrong:1999,Vinet:2006fj}.
For simplicity, we will not present its exact form here.
Instead, we give a detailed definition in~\eqref{eq:ResponseFunctionFullAppendix} of Appendix~\ref{app:geometry}.
Nevertheless, let us point out its main features.
Naively, the components of the response function are given by projections of the velocity perturbations of the nodes onto the interferometer arms,
\begin{equation}
    X(t) \sim \sum_{i,j=1}^3 \vec{n}_i \cdot \frac{\vec{v}_j(t-a_{ij}L/c)}{c} \, .
\label{eq:ResponseFunctionSchematic}
\end{equation}
Here, the $\vec{n}_i$ denote the unit vectors pointing between two nodes, labelled by the opposite side of the triangle, $\vec{v}_i$ is the velocity perturbation of the $i$-th node induced by the gravitational pull and $c$ is the speed of light.
The integer coefficients $a_{ij}$ take into account retardation effects along the different signal paths.
For more details on this notation see Appendix~\ref{app:geometry} and also~\cite{Armstrong:1999,Dhurandhar:2002zcl} for alternative response functions.

In fact, in some situations the detector response function above can be simplified (see, e.g.,~\cite{Vinet:2006fj}).
For instance, if the velocity perturbations of two nodes are negligible for all practical purposes, e.g., $\vec{v}_2 \approx \vec{v}_3 \approx 0$, it reduces to
\begin{equation}
	X(t) = -\vec{n}_1 \cdot \frac{\vec{v}_1(t) - \vec{v}_1(t - 4 L/c)}{c} \, ,
\label{eq:ResponseFunction}
\end{equation}
where $\vec{v}_1(t)$ is the velocity perturbation of a \emph{single} node and we have used that $\vec{n}_1 + \vec{n}_2 + \vec{n}_3 = 0$.
The latter condition is true, if all interferometer arms are of equal length.

Carefully note that this form of $X(t)$ is only an approximation of the detector response.
Eq.~\eqref{eq:ResponseFunction} is the dominant contribution to the exact response function, if the dark matter clump approaches the detector node very closely, or, in other words, if the impact parameter is smaller than the arm length of the interferometer, $D \lesssim L$.
This is the so-called \textit{close-approach} limit (see, e.g.,~\cite{Vinet:2006fj,Seto:2004zu}).
In contrast, if the impact parameter is much larger than the arm length (sometimes called the \textit{tidal} limit, see~\cite{Vinet:2006fj,Seto:2004zu}), $D \gg L$, the differential gravitational pull on the nodes can be qualitatively different from the pull on a single node.
To see this, imagine a simple situation where two nodes are aligned on an axis perpendicular to the trajectory of the dark matter clump, with their relative distance to each other being much smaller than their distance to the clump's trajectory.
In this scenario, the impact parameters of both satellites will differ by the arm length, i.e.~they read $D$ and $D + L$, respectively.
According to~\eqref{eq:ResponseFunctionSchematic}, the detector response will be proportional to the differential gravitational acceleration between the two nodes, with impact parameters and $D$ and $D+L$,
\begin{equation}
	X(t) \propto g(D) - g(D+L) \sim \frac{1}{D^2} - \frac{1}{\left(D+L\right)^2} \, .
\end{equation}
Hence, comparing both regimes, we observe the behavior
\begin{equation}
	X(t) \sim \begin{cases}
						\frac{1}{D^2} \left(1 - \frac{D^2}{L^2}\right) \, , & D \ll L \\
						\frac{L}{D^3} \, , & D \gg L
					\end{cases}
					\, .
\label{eq:ResponseRegimes}
\end{equation}
That is, the detector response in the tidal regime, $D \gg L$, falls off much faster with distance than in the close-approach limit, $D \ll L$.
In this regime, we therefore expect the close-approach approximation of the detector response to break down and one would need to take the exact response function into account.
Nevertheless, we will argue \textit{a posteriori} that, for the purpose of this section, we only have to consider the close-approach case, $D \lesssim L$, for events with a detectable signal-to-noise ratio.
Therefore, \eqref{eq:ResponseFunction} gives a reasonable approximation to the detector response.
We note however, that when the sensitivity of the experiment becomes better, i.e.~the noise is reduced, events with larger impact parameters will become detectable and the close-approach approximation might have to be reconsidered.

\bigskip

In practice, by means of~\eqref{eq:ResponseFunction} we can compute the response of the interferometer to an arbitrary velocity perturbation.
For concreteness, we have considered the velocity perturbation associated to a gravitational pull by a dark matter clump passing by a single LISA satellite in~\eqref{eq:DMacceleration} in a specific coordinate frame.
However, in principle, the dark matter clump can approach the satellite from any direction.
In order to account for this, we can equivalently choose an arbitrary orientation of the LISA experiment.
That is, we can parametrize the unit vector $\vec{n}_1$ in~\eqref{eq:ResponseFunction} by $\vec{n}_1 = \left( \sin\vartheta \cos\varphi, \sin\vartheta \sin\varphi, \cos\vartheta\right)$.
The angles $\vartheta$ and $\varphi$ essentially implement the arbitrary orientation of the detector plane relative to the dark matter clump.
For a detailed discussion of this, see Appendix~\ref{app:GeomSpheres}.
As we are interested in the signal power spectrum~\eqref{eq:PowerSpectrum}, it is then natural to consider the Fourier transform\footnote{In this work, we use the symmetric convention of the Fourier transform, $\tilde{f}(\omega) = (2\pi)^{-1/2} \int_\mathbb{R} \diff t \, f(t) \exp(i\omega t)$.} of the response function, $\tilde{X} (\omega)$.
In our scenario, the latter is given by (cf.~\cite{Vinet:2006fj}, adding suitable time delays which give the $L$ dependent factor)
\begin{equation}
\begin{split}
	\tilde{X} (\omega) =& \sqrt{\frac{2}{\pi}} \left(1 - \me^{4i \omega L / c}\right) \frac{GM}{cV^2} \sin\vartheta \\
	& \times \left[ K_0 \left(\frac{D\omega}{V}\right) \sin\varphi - i K_1 \left(\frac{D\omega}{V}\right) \cos\varphi \right] \, ,
\end{split}
\end{equation}
where $K_i$ is the $i$-th modified Bessel function of the second kind.
In principle, we could now obtain the signal power spectrum $P(\omega)$ by squaring this expression.
Obviously, the spectrum would then depend on the orientation of the LISA detector relative to the trajectory of the dark matter clump through the angles $\vartheta$ and $\varphi$.
As a simple approximation it is reasonable to assume that the dark matter clumps generically move in a random direction.
That means, in principle, any value of $\vartheta$ and $\varphi$ is equally likely.
To account for this, we can assume a uniform distribution for both, of which we then take the average (see Appendix~\ref{app:GeomSpheres} for details).
Before proceeding we note, however, that a uniform average is only a rough approximation.
It assumes that the local dark matter distribution is isotropic. 
While this may be true in an isotropic reference frame where the detector is at rest, the Sun, together with the experiment, is moving through the dark matter halo at a constant velocity, thereby imposing a \emph{preferred} direction on the system.
Therefore, strictly speaking, the distribution of the angles parametrizing the relative orientation between the dark matter and the detector plane should not be uniform.
This, however, is somewhat ameliorated by the motion of LISA itself (see also~\cite{Vinet:2006fj}), which changes direction along its orbit around the sun (cf., e.g.,~\cite{Danzmann2017LISA}).
While certainly not all directions are attained with equal probability, it nevertheless amounts to at least a partial averaging.
For a more detailed discussion of this see~\ref{app:dmvelocity}.

\begin{figure}[t]
\centering
	\includegraphics[width=0.85\columnwidth]{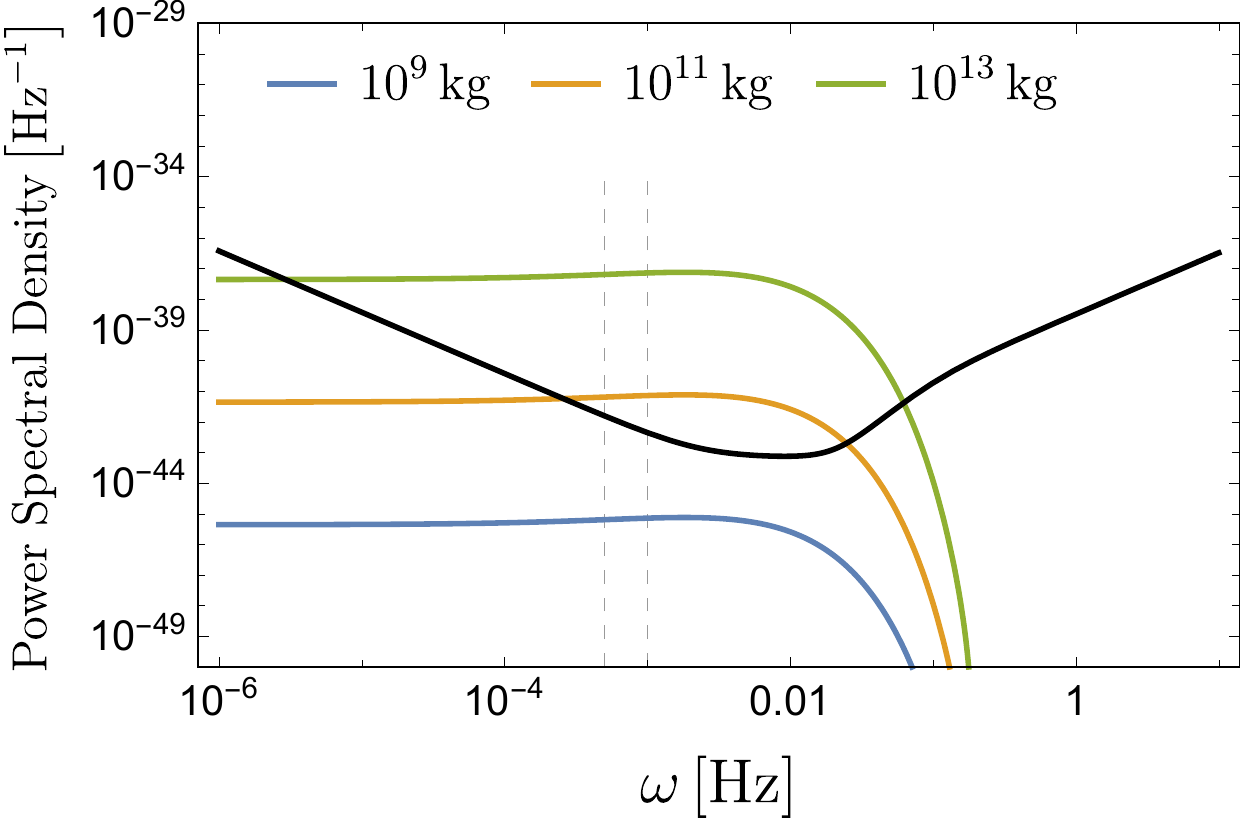}
	\caption{Angular-averaged signal power spectral density of dark matter clumps compared to the experimental sensitivity of LISA. The signal is shown in color, illustrating different masses of the clumps, $M_{\mathrm{DM}}$. The sensitivity of LISA, i.e.~the noise spectrum~\eqref{eq:NoisePowerSpectrum}, is shown in black. The dashed lines illustrate possible choices of cutoff frequencies $\omega_c$ for the estimation of the signal-to-noise ratio. Here, we consider events with an impact parameter of $D=50,000 \, \mathrm{km}$. We furthermore fix $\rho_{\mathrm{DM}} \approx 0.39 \, \mathrm{GeV/cm^3}$ and assume a velocity distribution of Maxwell-Boltzmann type for the dark matter clumps with root mean square $v_{\mathrm{rms}} \approx 270 \, \mathrm{km/s}$ in combination with the LISA experiment moving through the galaxy at $v_{\odot} \approx 220 \, \mathrm{km/s}$.}
\label{fig:DM-Signal}
\end{figure}

Finally, with this caveat in mind, we arrive at the angular-averaged signal power spectrum associated to the gravitational pull of a dark matter clump on an interferometer node~\cite{Vinet:2006fj},
\begin{equation}
\begin{split}
	P(\omega) = \avg{\abs{\tilde{X} (\omega)}^2} = & \frac{8}{3\pi} \left(\frac{GM}{cV^2}\right)^2 \sin^2 \left( 2 \omega L / c\right) \\
	& \times \left[K_0^2 \left(\frac{D\omega}{V}\right) + K_1^2 \left(\frac{D\omega}{V}\right) \right] \, .
\end{split}
\label{eq:signalpowerspectrum_angleavg}
\end{equation}
For a few examples of the typical shape of $P(\omega)$ we refer the reader to Fig.~\ref{fig:DM-Signal}.
Note that there, for reasons that will become clear momentarily, we show the signal power spectral density, $4 \omega P(\omega)$.

Obviously, the signal power spectrum is only useful for an experimental test, if the desired signal can be distinguished from the background noise that the experiment is subject to.
In general, at a gravitational wave interferometer the background noise is characterized by a noise power spectrum, commonly defined by $\avg{\tilde{n}(f) \tilde{n}^{\ast}(f^{\prime})} = \frac{1}{2} \delta(f-f^{\prime}) S_n(f)$ (for a comprehensive review see, e.g.,~\cite{Moore:2014lga}).
For future experiments such as LISA this is not yet completely settled.
In principle, different estimates can lead to quite different results for the detection rate.
For concreteness, we will use a recent estimate~\cite{Cornish:2018dyw} as our benchmark,
\begin{equation}
\begin{split}
	S_n(\omega) =& \frac{1}{2\pi} \left( \frac{2 \left(\omega L / c\right)^2}{1 + \left(\omega L / c\right)^2} \right)^2 \frac{10}{3L^2} \\
	&\times \left[ B_1 (\omega) +\frac{2 + 2\cos^2 \left( \omega L/c \right)}{\omega^4}  B_2 (\omega) \right] \\
	&\times \left[ 1+ \frac{6}{10} \left( \frac{\omega L}{c} \right)^2 \right] \, ,
\end{split}
\label{eq:NoisePowerSpectrum}
\end{equation}
with
\begin{align}
	B_1(\omega) &= \left( 1.5 \times 10^{-11} \, \mathrm{m} \right)^2\left(1 + \left( 2\pi \frac{2 \, \mathrm{mHz}}{\omega} \right)^4 \right) \, \mathrm{Hz^{-1}} \, , \\
\begin{split}
	B_2(\omega) &= \left( 3 \times 10^{-15} \, \mathrm{ms^{-2}}\right)^2 \left(1 + \left( 2\pi \frac{0.4 \, \mathrm{mHz}}{\omega}\right)^2\right) \\ &\times \left( 1 + \left(\frac{1}{2\pi} \frac{\omega}{8 \, \mathrm{mHz}}\right)^4\right) \, \mathrm{Hz^{-1}} \, .
\end{split}
\end{align}
Note that, here, we have added an additional factor of $4 (\omega L / c)^4/(1 + (\omega L / c)^2)^2$ compared to~\cite{Cornish:2018dyw}.
This essentially acts as an estimate of a transfer function to convert the original strain spectrum to the equivalent of our signal spectrum $\tilde{X}$\footnote{For example, following~\cite{Armstrong:1999}, this can be seen explicitly by evaluating a strain signal of a gravitational wave in the channel $X(t)$, yielding a transfer function proportional to $4 \sin^4 (\omega L /c)$. In this case our factor corresponds to the envelope of this conversion factor. Note that this approximation is correct up to a possible constant factor of order $\mathcal{O}(1-10)$.}.
Furthermore, we have divided by $2\pi$ in order to match our (symmetric) convention of the Fourier transform to the convention typically used in signal processing.
We further remark that other choices of a noise power spectrum might also be reasonable, for example, the ones presented in~\cite{Bender1998LISA,Danzmann2017LISA}.

Finally, as a measure for distinguishing a signal from the background noise, we can define the signal-to-noise ratio by comparing the signal- to the noise power spectrum over the range of all frequencies (see, e.g.,~\cite{Saulson:1995zi}),
\begin{equation}
\begin{split}
	\mathrm{SNR} &= \left(4 \int_0^\infty \diff \omega \, \frac{P(\omega)}{S_n(\omega)} \right)^\frac{1}{2} \\
	&= \left( \int_{-\infty}^{\infty} \diff \left( \log \omega \right) \, \frac{4\omega P(\omega)}{S_n(\omega)} \right)^\frac{1}{2} \, .
\end{split}
\label{eq:SNR}
\end{equation}
Here, one factor of 2 arises from the definition of the noise power spectral density given above.
A second factor of 2 reflects the fact that we are considering single-sided power spectra only~\cite{Saulson:1995zi,Vinet:2006fj,Moore:2014lga}.

The form on the very right hand side of the equation is particularly useful for a quick estimation of the signal-to-noise ratio from the usual logarithmic plots of the sensitivity, as it is obtained from an integral of the logarithm over a dimensionless ratio between the signal, $4\omega P(\omega)$, and the noise, $S_n(\omega)$, power spectral density.
Before continuing, let us remark that the signal power spectrum $P$ given in~\eqref{eq:signalpowerspectrum_angleavg} remains constant with decreasing frequency due to the constant velocity of the satellite in the asymptotic future.
Nevertheless, this does not lead to an infinite signal-to-noise ratio when normalizing to our choice of the noise power spectrum, $S_n$.
Instead, it remains finite, even when integrated over all frequencies.
As was already noted in~\cite{Vinet:2006fj}, alternative noise power spectra, however, might not share this feature and hence require the introduction of an experiment-specific lower frequency cut-off.
For LISA imposing a cutoff $\omega_c \sim 10^{-4} \, \mathrm{Hz}$ due to experimental limitations provides a relatively conservative estimate of the lower end of the frequency band that the detector is sensitive to (see, e.g.,~\cite{Vinet:2006fj}).
However, in this paper we use the close-approach approximation of dark matter clump encounters with the interferometer.
This is valid for impact parameters smaller than the size of the experiment, $D \lesssim L$.
Therefore, we will use an even more conservative cut-off, that is essentially determined by the characteristic time of flight of a dark matter clump through the detector volume, $\omega_c \sim 2\pi V/L \sim 10^{-3} \, \mathrm{Hz}$.

In Fig.~\ref{fig:DM-Signal} we show the angular-averaged signal power spectral density, $4\omega P(\omega)$ with $P(\omega)$ given in~\eqref{eq:signalpowerspectrum_angleavg}, associated to dark matter clumps traversing the detector volume and compare it to the experimental sensitivity of LISA.
As an example, we consider events with an impact parameter of $D=50,000 \, \mathrm{km}$.
The coloured lines correspond to the signal, while the black solid line shows the noise power spectrum $S_n(\omega)$ given in~\eqref{eq:NoisePowerSpectrum}, which essentially determines the overall experimental sensitivity of the detector.
As indicated earlier, the signal-to-noise ratio is the logarithimic integral over the ratio of the two plotted quantities and we can roughly read this figure in such way, that we have a chance to distinguish a signal from the detector noise, whenever the coloured signal exceeds the black noise spectrum for a sufficient range of frequencies.
Also note that the signal power spectral density shown here, is uniformly averaged over the angles parametrizing the relative orientation between the interferometer and the trajectory of the dark matter clump (cf.~Appendix~\ref{app:GeomSpheres}).
As explained above, this approximates the typical size of the signal.
The signal shape and strength of individual events will be different.
For instance, in extreme cases, we expect a different signal from a clump with normal incidence to the detector plane as from one that traverses the detector volume almost parallel to it.
Furthermore, as we have already pointed out before, the uniform average can only be seen as an approximation.
This is because there is preferred direction given by the Sun, together with the detector, moving through the dark matter halo at a constant velocity (see Appendix~\ref{app:dmvelocity}) as well as the rotation of the detector itself not covering all angles equally.

In general, the signal power spectrum receives large contributions from low frequencies, while it quickly drops in the high frequency regime.
As pointed out earlier, the constant value at low frequencies is due the constant velocity component of the LISA satellite in the asymptotic future.
Consequently, the sensitivity may benefit from background noise that is reduced in the low frequency tail of the spectrum.
This, however, will also require going beyond the close approach approximation.
Keeping this caveat in mind, below we investigate the benefits of improving at low frequencies by considering two different low-frequency cutoffs.

\bigskip

Having reviewed all necessary aspects of distinguishing a possible signal induced by a localized clump of dark matter traversing the detector volume of LISA from background noise, let us now quantify the discovery potential for these clumps inside the dark matter halo of our Galaxy.
The mass of the dark matter clumps, $M_{\mathrm{DM}}$, determines the characteristic distance between them, $d \sim \left( M_{\mathrm{DM}} / \rho_{\mathrm{DM}} \right)^{1/3}$.
Therefore, it controls the average rate of encounters of a clump with one of the satellites at or below a given impact parameter $D$,
\begin{equation}
	\dot{\eta} = \pi D^2 \Phi_{\mathrm{DM}} \, ,
\end{equation}
where $\Phi_{\mathrm{DM}}$ is the effective dark matter flux at velocity $v_{\mathrm{DM}}$, given by $\Phi_{\mathrm{DM}} \sim \rho_{\mathrm{DM}} v_{\mathrm{DM}} / M_{\mathrm{DM}}$.
That is, naively, $\dot{\eta}$ is the rate at which, on average, a dark matter clump of mass $M_{\mathrm{DM}}$ and velocity $v_{\mathrm{DM}}$ passes through a surface of radius $D$.
Therefore, applied to our scenario, we can use it to estimate the rate at which we expect a dark matter clump to induce a signal in the interferometer.
Note, however, that using the average signal strength in the calculation of the required impact parameter and therefore $\dot{\eta}$ is only an approximation.
In general, clumps passing by at a suitable angle may already give rise to a signal at somewhat larger distances than indicated by the average signal strength and, similarly, closer encounters are needed for other angles. 
However, we expect that this simplistic treatment nevertheless captures the effect reasonably well.
Furthermore, we assume the velocity distribution of the dark matter clumps inside the halo to be a (simplified) superposition of the Sun moving through the Galaxy at $v_{\odot} \approx 220 \, \mathrm{km/s}$~\cite{GalacticConstants1986} and the dark matter velocity having a uniformly random direction\footnote{Indeed, this is already implemented by the uniform average of the solid angle in~\eqref{eq:signalpowerspectrum_angleavg}. While this is still an approximation, we expect to obtain results to a reasonable accuracy. We give a discussion of this in Appendix~\ref{app:dmvelocity}.} while its magnitude is Maxwell-Boltzmann distributed with a root mean square of $v_{\mathrm{rms}} = \sqrt{3/2} v_{\odot} \approx 270 \, \mathrm{km/s}$ (see, e.g.,~\cite{Kamionkowski:1997xg}).
For a discussion of possible caveats of this approximation and how it is implemented in practice, see Appendix~\ref{app:dmvelocity}.
In addition, we fix the dark matter energy density to $\rho_{\mathrm{DM}} \approx 0.39 \, \mathrm{GeV/cm^3}$~\cite{Catena:2009mf}.
Note that these values can come with relative uncertainties of up to $25 \, \%$, which could alter our results by a similar amount.
We remark, however, that the approximations we employ as well as the differences in the noise power spectra probably cause larger uncertainties.

\bigskip

\begin{figure}[t]
\centering
	\includegraphics[width=0.85\columnwidth]{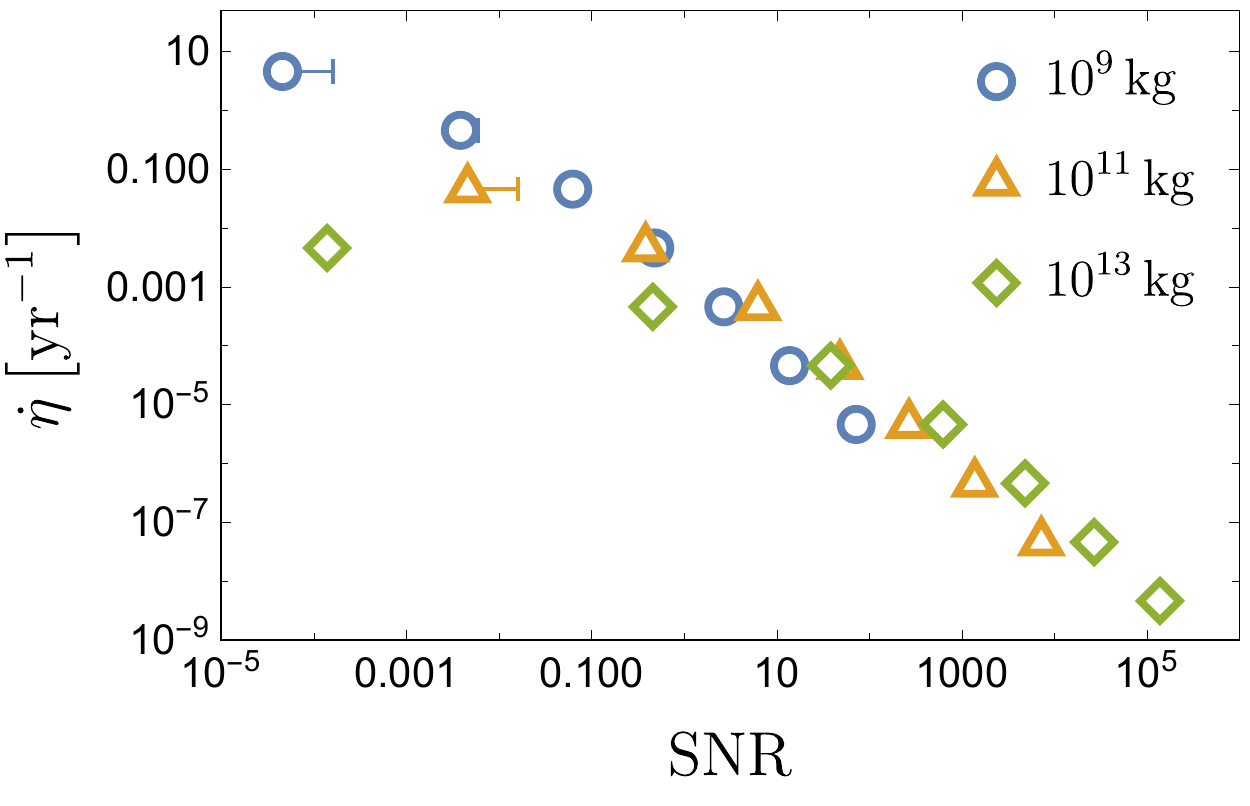}
	\includegraphics[width=0.95\columnwidth]{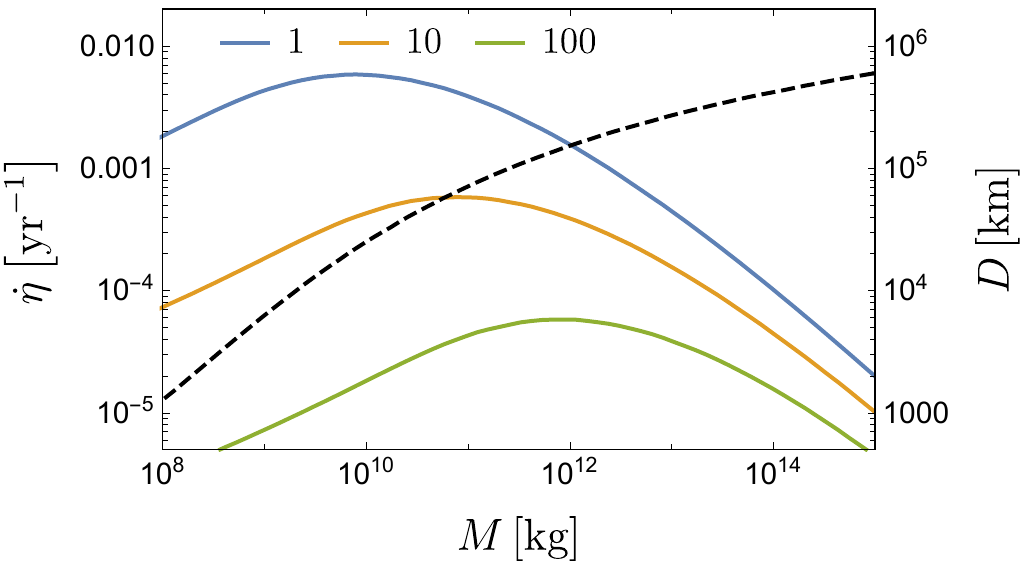}
	\caption{Average gravitational interaction rate of localized dark matter clumps with the LISA nodes as a function of the signal-to-noise ratio (top) and the mass of the dark matter clumps (bottom). In the top panel the colors illustrate different masses of the clumps, while in the bottom panel they denote a constant signal-to-noise ratio.  The bars in the top panel illustrate the dependence of the signal-to-noise ratio on the low-frequency cutoff described in the main text. The black dashed line in the bottom panel indicates the impact parameter at which $\mathrm{SNR} \gtrsim 1$. As long as this is smaller than the size of LISA we expect the close-approach approximation to be valid. We use the same baseline parameters for the dark matter density and velocity as in Fig.~\ref{fig:DM-Signal}.}
\label{fig:DM-Eta}
\end{figure}

In Fig.~\ref{fig:DM-Eta} we show the average gravitational interaction rate as a parametric function of the signal-to-noise ratio as well as the mass of the dark matter clumps.
As pointed out before, here, we only take impact parameters into account that are smaller than the arm length of LISA, $D \lesssim L$, i.e.~we consider the close-approach regime.
This is indicated by the black dashed line in the bottom panel (with the appropriate scale being on the right hand side), which illustrates typical impact parameters at which the signal-to-noise ratio exceeds one, $\mathrm{SNR} \gtrsim 1$, for a given mass of the dark matter clump.
We can see that this is always smaller than the typical size of LISA and therefore the close-approach approximation of the detector response function~\eqref{eq:ResponseFunction} is valid in the mass regime we show here.
This approximation has to be reconsidered, if the sensitivity of the experiment is improved.
In this case, dark matter clumps at larger impact parameters can still be detected.
The impact parameter shown in Fig.~\ref{fig:DM-Eta} also gives an upper bound on the size of the dark matter clumps for which our treatment of the clumps as point-like is valid (most of the signal rate originates from the largest detectable impact parameters).

In general, we observe that with increasing mass of the dark matter clumps, the signal-to-noise ratio is enhanced.
However, at the same time, the event rate of encounters is reduced due to the reduced effective dark matter flux, $\Phi_{\mathrm{DM}} \sim M_{\mathrm{DM}}^{-1}$.
That is, at higher masses there is a balance between the increased signal-to-noise ratio and the reduced interaction rate.
In particular, we find an optimal detection potential that passes a minimal detection threshold, $\mathrm{SNR}\gtrsim 1$, for dark matter clumps of mass $M_{\mathrm{DM}} \approx 10^{10} \, \mathrm{kg}$ which could be observed at LISA approximately every 200 years on average.
However, we note that this threshold of a signal-to-noise ratio, $\mathrm{SNR}\gtrsim 1$, is rather optimistic.
In practice, as the signal itself has to be distinguished from other sources, the detection threshold may be significantly higher, of the order of $\mathrm{SNR}\gtrsim 10$ (e.g.~this is on the lower end of the typical signals considered in the Mock LISA Data Challenges~\cite{Arnaud:2006gm,MockLISADataChallengeTaskForce:2009wir}).
As shown in Fig.~\ref{fig:DM-Eta}, this will reduce the average detection rate by an order of magnitude while shifting the optimal detection potential to dark matter clumps of higher masses.

Unfortunately, the sensitivity is not yet on a desirable level for a near future discovery potential.
Let us note, however, that the latter value has to be understood as a relatively conservative estimate, which could also be significantly higher.
As already mentioned, the signal-to-noise ratio depends on the noise power spectrum which features a comparably large uncertainty.
For the relatively broad signal spectra associated to the dark matter clumps, in particular the low frequency tails of the spectrum may be relevant.
Therefore, the signal-to-noise ratio may vary significantly, if the overall noise spectrum was smaller, to some degree also at low frequencies.
We try to illustrate this through the bars in the top panel of Fig.~\ref{fig:DM-Eta}.
For these, we push the close-approach approximation closer to the boundary of its validity by introducing a low-frequency cutoff of $\omega_c \approx 5 \times 10^{-4} \, \mathrm{Hz}$ as compared to the one imposed by the close-approach approximation, $\omega_c \approx 10^{-3} \, \mathrm{Hz}$.
Let us stress, however, that this really has to be understood only as a very first estimate of potential improvements.
Finally, we note that, in light of the very rough nature of our estimates, the results presented Fig.~\ref{fig:DM-Eta} are reasonably in line with but perhaps somewhat more pessimistic than earlier works, in particular~\cite{Hall:2016usm}, with regards to the detection of dark matter clumps LISA.

While the obtained rate is still rather low, it nevertheless brings us much closer to a desired level, so that we can hope that further improvements both in the detector as well as in the analysis might allow for a detection in a reasonable time frame.
Indeed, one such improvement could be in the analysis.
For example, already the authors of~\cite{Adams:2004pk} noted that the dark matter clump interaction is inelastic and therefore differs from the elastic interaction of a gravitational wave exploited in current detection strategies.
Further improvements are expected from a more detailed analysis of the time structure of potential signals as discussed in~\cite{Dror:2019twh}.

\section{Topological Defects}
\label{sec:topologicalobjects}

In this section, we now want to go beyond dark matter clumps and investigate the detection of structures such as cosmic strings and domain walls\footnote{In principle, we could also consider monopoles.
However, the gravitational potential of a topological monopole is usually simply that of a point-source or particle~\cite{Barriola:1989hx,Harari:1990cz} and therefore already included in our discussion of clumps (if such super-massive monopoles were to exist).}.
As already mentioned in the introduction, these topological defects might have been produced in the early Universe~\cite{Kibble:1980mv,Vilenkin:1984ib,Press:1989yh}, but their use as dark matter requires a non-trivial behavior, e.g.~interacting networks, of such objects.
We do not address how such a network is formed or how it can be made to satisfy the constraints imposed by the properties of dark matter (or dark energy) but simply assume its presence with a given density. 
This is in the spirit of and follows the completely phenomenological approach to study their detection, also pursued in~\cite{Pospelov:2012mt,Derevianko:2013oaa,Motohashi:2013kka,Stadnik:2014tta,Stadnik:2014ala,Stadnik:2014cea,Jaeckel:2016jlh,Roberts:2017hla,Grote:2019uvn,Masia-Roig:2019hsy,McNally:2019lcg} in the context of various different detection techniques.
We briefly note that our discussion also applies to localized structures of ``ordinary'' dark matter with a string-like or domain wall-like geometry, an example of which are the string-like axion streams that were already investigated with regard to LISA in~\cite{Dokuchaev:2017psd}.

The gravitational properties of topological objects can significantly differ from those of non-relativistic matter.
Cosmologically, this is due to a different equation of state, relating the pressure and the energy density of the cosmic fluid, $w = p / \rho$.
This then results in a typical scaling behavior $\rho\sim a^{-3(1+w)}$ of the average energy density on cosmological scales.

For our signals we are interested also in the behavior in the vicinity of individual objects. Very naively, this can be obtained from Poisson's equation for the Newtonian gravitational potential.
For a \emph{fluid} this is sourced by the combination of pressure and energy $(\rho + 3 p)$ (see, e.g.,~\cite{Peebles:1994xt}), and the gravitational potential satisfies
\begin{equation}
	\Laplace \phi = 4 \pi G \left(1 + 3w\right) \rho \, ,
\label{eq:PoissonNewton}
\end{equation}
i.e.~it explicitly depends on the equation of state of the source.

However, it is not obvious that the gravitational field of topological objects such as strings and domain walls behaves \emph{locally} as suggested by their global equation of state.
That said, in good approximation this is nevertheless true for static strings as well as domain walls, as shown by the results of~\cite{Vilenkin:1981zs,Vilenkin:1984hy} and discussed below for each case in the respective subsection.
To model the local field of more complicated networks of topological structures we therefore simply assume that it can be approximated by Eq.~\eqref{eq:PoissonNewton} with the equation of state parameter given by its cosmological value.

The equation of state as well as their dimension for cosmic vacuum strings and domain walls leads to significantly different gravitational fields sourced by these objects compared to non-relativistic clumps of matter.
In the following subsections, we want to discuss their gravitational properties and the related imprints they might leave at LISA.

\subsection{Cosmic strings}
\label{subsec:CosmicStrings}

In order to obtain the gravitational potential of a cosmic string, we can solve~\eqref{eq:PoissonNewton} in a cylindrically symmetric space for an energy density that is distributed along an infinite string, e.g.~$\rho = \mu \delta (x) \delta (y)$.
Here, $\mu$ is the tension of the string, i.e.~the energy stored per unit length.
This yields a gravitational potential that grows logarithmically with distance, $\phi \sim \log \left( r_0/r \right)$, and sources the gravitational field
\begin{equation}
	\vec{g}(\vec{r}) = -2  (1+3w) \frac{G\mu}{r} \vec{e}_r \, ,
\label{eq:CSGravitationalField}
\end{equation}
where $r$ denotes the radial distance to the string.

In general, we can directly use~\eqref{eq:CSGravitationalField} to obtain the signal strength at LISA. However, before we proceed, let us mention a noteworthy special case.
If we consider a single \emph{static} vacuum string, the equation of state is given by $w=-1/3$, and its energy density dilutes as $\rho_{\mathrm{S}} \sim a^{-2}$~\cite{Kolb:1990vq}.
At the same time, this implies that its gravitational field vanishes, indicating that a static cosmic string does not couple to matter~\cite{Vilenkin:1981zs,Hiscock:1985uc}.
Nevertheless, this picture can change, if the vacuum string starts to be dynamical.
For instance, a string moving at a velocity $\beta=v/c$ has a modified equation of state, $w = 2/3\beta^2 - 1/3$~\cite{Kolb:1990vq}.
Going beyond such scenario, the situation can deviate from this even more, if multiple strings are considered.
In particular, if the strings interact with each other to form a network, the corresponding equation of state can drastically change.
Therefore it is possible that they can contribute to the dark matter or dark energy component of the Universe (see, e.g.,~\cite{Bucher:1998mh}).

The naive reasoning from above suggests, that we would not expect any gravitational pull on the interferometer by a static vacuum string at all.
However, it was shown that, due to the globally non-trivial conical spacetime geometry sourced by the string, it will still attract massive objects around it~\cite{Vachaspati:1990am}.
In fact, the gravitational acceleration of a mass $m$ in the vicinity of the string is given by~\cite{Vachaspati:1990am}
\begin{equation}
	\vec{g}(\vec{r}) = - 8 \pi \kappa G \frac{ (G \mu /c^2) m}{r^2} \vec{e}_r \, ,
\label{eq:CS_gravfield}
\end{equation}
where $\kappa \approx 1/32$ for small $G \mu / c^2$.
In this case, the gravitational field falls of as $g \sim r^{-2}$ in contrast to the $r^{-1}$ asymptotics of the general configuration~\eqref{eq:CSGravitationalField}.
Naively, this can be understood as follows.
An infinite, straight and static cosmic string sources a conical spacetime geometry that can lead to double copies, i.e.~mirror images, of nearby objects~\cite{Vilenkin:1981zs,Hiscock:1985uc}.
In this sense, \eqref{eq:CS_gravfield} can be understood as the gravitational field sourced by the effective mass, $(G \mu /c^2) m$.
Hence, we can interpret the gravitational field of the string as similar to the usual attractive force between two massive objects.
Cosmological considerations, such as the isotropy of the cosmic microwave background, provide bounds on the dimensionless string tension, which are typically of the order $G \mu / c^2 \lesssim 10^{-6}$ (see, e.g.,~\cite{Hindmarsh:1994re,Vilenkin:2000jqa}).
Therefore, we expect a tiny acceleration of a test mass in the gravitational field of a static cosmic string.

In the following, we will distinguish between signals due to the gravitational pull on the interferometer generated by a static vacuum string and by an interacting network of cosmic strings that accounts for dark matter or dark energy.
The former will be characterized by the gravitational field~\eqref{eq:CS_gravfield}, while the latter is given by~\eqref{eq:CSGravitationalField} with a general equation of state.

\subsubsection*{Signal power spectrum}

Given the gravitational field of a cosmic string, we can proceed analogously to the case of dark matter clumps and determine the gravitational acceleration that each node of the interferometer experiences in the vicinity of a string.
For concreteness, let us choose a coordinate frame, in which the (infinite) string is parallel to the $z$-axis and the LISA satellite located at the origin is, initially at $t_0=0$, at a minimum distance $D$ to the string.
We furthermore assume that the string is uniformly moving at velocity $V$.
Due to the additional internal orientation of the string as compared to spherical clumps, in the frame where the string motion is confined to the $yz$-plane, its velocity can have a component in the $y$- as well as the $z$-direction, i.e.~$V_y = V \sin\theta$ and $V_z = V \cos\theta$, respectively.
In other words, the string has an additional inclination angle when approaching the interferometer node.
For a schematic illustration see the bottom panel of Fig.~\ref{fig:LISAobjects} in Appendix~\ref{app:geometry}.
In this reference frame, in the gravitational field of a \emph{static} cosmic string~\eqref{eq:CS_gravfield}, a test mass is subject to an acceleration of
\begin{equation}
	\vec{g}(t) = \frac{8 \pi \kappa G \left(G \mu / c^2\right) m}{D^2\left(1 + \left(\frac{Vt}{D}\right)^2 \sin^2\theta\right)^{3/2}}
	\begin{pmatrix}
													1 \\ \frac{Vt}{D} \sin\theta \\ 0
												\end{pmatrix} \, ,
\label{eq:GravFieldCSStatic}
\end{equation}
while in the field of an interacting string network~\eqref{eq:CSGravitationalField} it reads
\begin{equation}
	\vec{g}(t) = \frac{2 G \mu (1+3w)}{D\left(1 + \left(\frac{Vt}{D}\right)^2 \sin^2\theta \right)}
	\begin{pmatrix}
													1 \\ \frac{Vt}{D} \sin\theta \\ 0
												\end{pmatrix} \, .
\label{eq:GravFieldCS}
\end{equation}
Note that in both expressions only the velocity component $V_y = V \sin\theta$ appears.
This is because only the radial distance to the string, in this reference frame determined by the $x$- and $y$-coordinate, enters the gravitational field.

Along the lines of our discussion in Section~\ref{sec:dmclumps}, in order to obtain the frequency spectrum of the detector response, we now need to determine the Fourier transform of the velocity perturbations associated to the accelerations of the detector nodes.
As we will again argue \emph{a posteriori}, we consider the regime where the strings traverse the detector volume in the close vicinity of a single node, i.e.~the close-approach limit, $D \ll L$.
Therefore, the detector response can be approximated by~\eqref{eq:ResponseFunction}, which in the frequency domain reads $\tilde{X}(\omega) \approx -\left(1 - \exp(-4i\omega L /c) \right) \vec{n}_1 \cdot \tilde{\vec{v}}(\omega)/c$.

In addition to the orientation of the string and its motion relative to the node of the interferometer, the arbitrary orientation of the detector plane has yet to be implemented.
In the close-approach approximation, we can take this into account by parametrizing $\vec{n}_1$ accordingly, $\vec{n}_1 = \left( \sin\vartheta \cos\varphi, \sin\vartheta \sin\varphi, \cos\vartheta \right)$.
For a detailed discussion of this see Appendix~\ref{app:GeomStrings}.

By squaring the Fourier transform of the response function, $\absl{\tilde{X}(\omega)}^2$, we obtain the signal power spectrum induced by a cosmic string in the vicinity of the interferometer.
At this point, the latter still depends on the relative orientation between the string and the detector plane, parametrized by $\vartheta$ and $\varphi$.
As already noted for the case of clumps, we assume that any orientation occurs equally likely (a discussion of the validity of this simplistic approximation and additional details are given in Appendix~\ref{app:dmvelocity}).
Therefore, we take the uniform average over both (cf.~Appendix~\ref{app:GeomStrings}).
Finally, the angular-averaged signal induced by the gravitational pull of a static cosmic string is given by
\begin{equation}
\begin{split}
	P(\omega) =& \frac{512 \pi}{3} \kappa^2 \left( \frac{G \left(G\mu/c^2\right) m}{c V^2 \omega}\right)^2 \sin^2 \left( 2 \omega L/c\right) \\
		&\times G_{3,0}^{1,3} \left( \ved{\frac{1}{2}}{-\frac{3}{2},0,1} \bigg| \left(\frac{D\omega}{V}\right)^2 \right) \, ,
\end{split}
\label{eq:CS-SignalPowerSpectrumStatic}
\end{equation}
where $G$ is the Meijer $G$-function.
For an interacting network of cosmic strings with an arbitrary equation of state, it reads
\begin{equation}
	P(\omega) = \frac{32}{3} \left(1+3w\right)^2 \left( \frac{G \mu}{c V \omega} \right)^2 \sin^2 \left( 2 \omega L/c\right) K_1 \left( \frac{2D\omega}{V} \right) \, .
\label{eq:CS-SignalPowerSpectrum}
\end{equation}
Here, $K_1$ denotes the first modified Bessel function of the second kind.

\begin{figure}[t]
\centering
	\includegraphics[width=0.85\columnwidth]{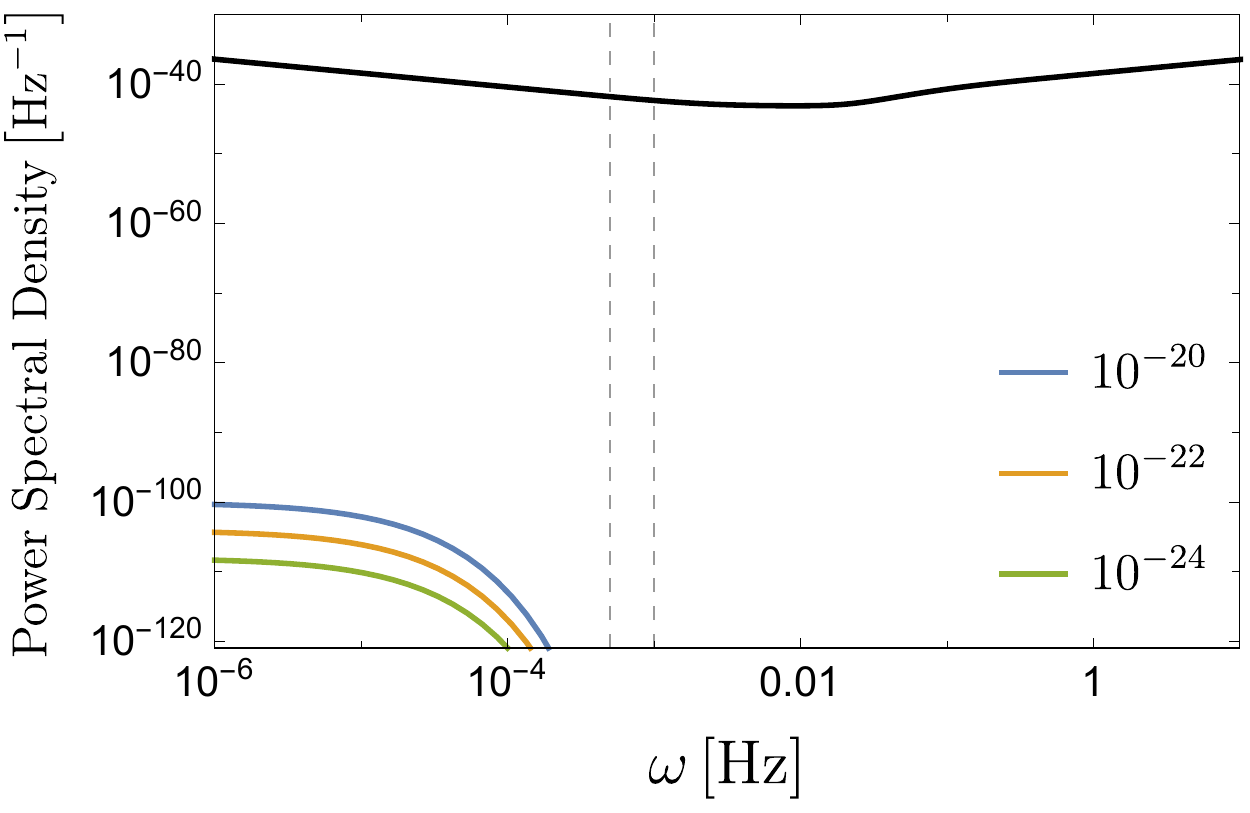}
	\includegraphics[width=0.85\columnwidth]{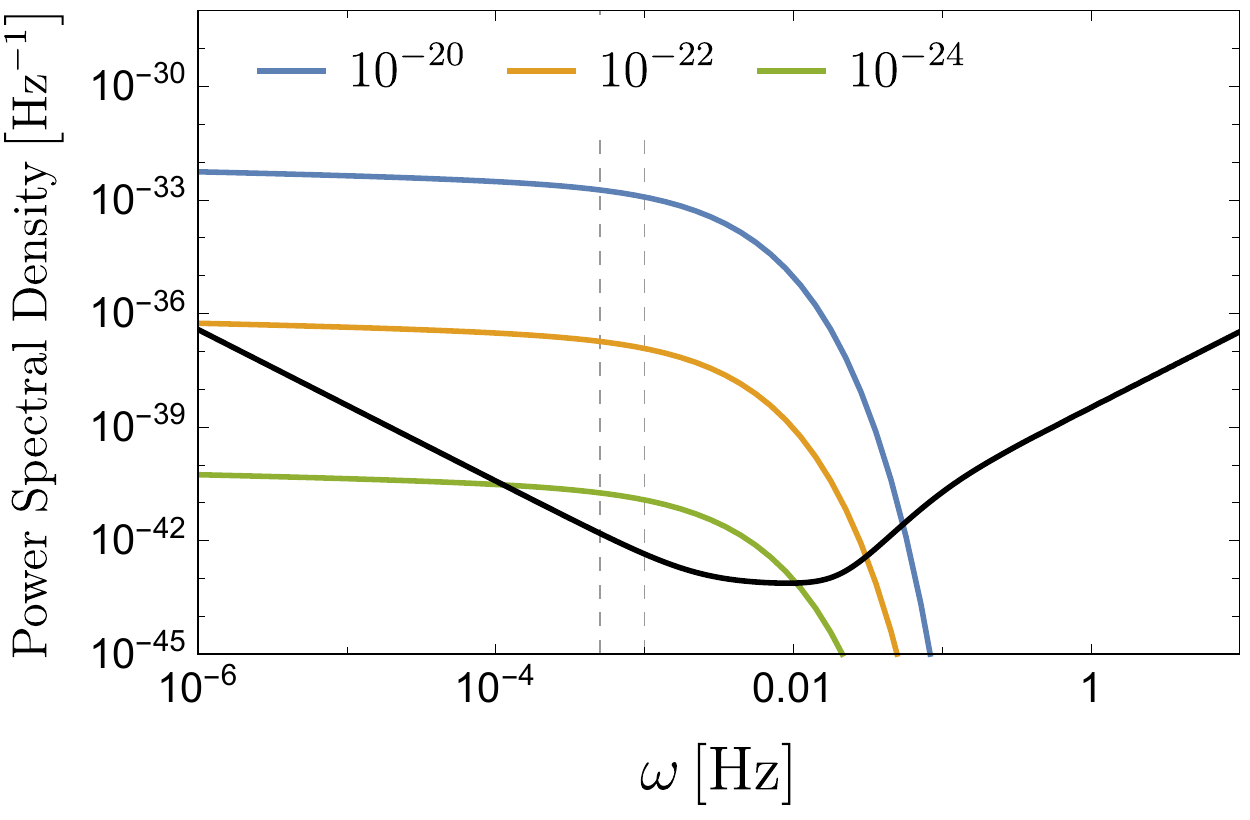}
	\caption{Angular-averaged signal power spectral density due to a static cosmic string (top) and an interacting network of cosmic strings (bottom) compared to the experimental sensitivity of LISA. The colors denote different values of the dimensionless string tension, $G\mu/c^2$. Here, we have chosen an impact parameter of $D=100,000 \, \mathrm{km}$. Therefore, as the strings are taken to account for dark matter, these signal events are very rare and occur every 10,000,000 (blue) to 1,000 (green) years on average. The black lines correspond to the sensitivity of LISA and the vertical dashed lines to two different values of the used low-frequency cutoff. In both panels we have assumed a Maxwell-Boltzmann distribution for the velocity of the strings with root mean square $v_{\mathrm{rms}} \approx 270 \, \mathrm{km/s}$ in combination with the LISA experiment moving through the galaxy at $v_{\odot} \approx 220 \, \mathrm{km/s}$.}
\label{fig:CS-Signal1}
\end{figure}

In the panels of Fig.~\ref{fig:CS-Signal1}, we illustrate an example of the angular-averaged signal power spectral densities induced by a static cosmic string (top) and an interacting string network with $w=0$ (bottom) and compare it to the experimental sensitivity of LISA.
In particular, we show different values of the string tension and consider, as an example, events with a fixed impact parameter of $D = 100,000 \, \mathrm{km}$.
Moreover, we have assumed a (simplified) superposition of a Maxwell-Boltzmann distribution for the velocity of the strings with root mean square $v_{\mathrm{rms}} \approx 270 \, \mathrm{km/s}$ in combination with the LISA experiment moving through the galaxy at $v_{\odot} \approx 220 \, \mathrm{km/s}$.
We present the practical implementation of this in Appendix~\ref{app:dmvelocity}.

Similar to the dark matter clumps, we find that the signal receives large contributions from low frequencies, while it quickly drops in the high frequency regime.
The signal induced by a static cosmic string is orders of magnitude smaller compared to the signal caused by interacting cosmic strings with $w=0$ as appropriate for the dark matter component of the Universe.
This is due to the fact that the gravitational perturbation of the former is sourced by the tiny mirror mass, $(G \mu / c^2) m$.
In fact, the suppression already indicates that only an interacting network of cosmic strings is within experimental reach of LISA.

\begin{figure}[t]
\centering
	\includegraphics[width=0.85\columnwidth]{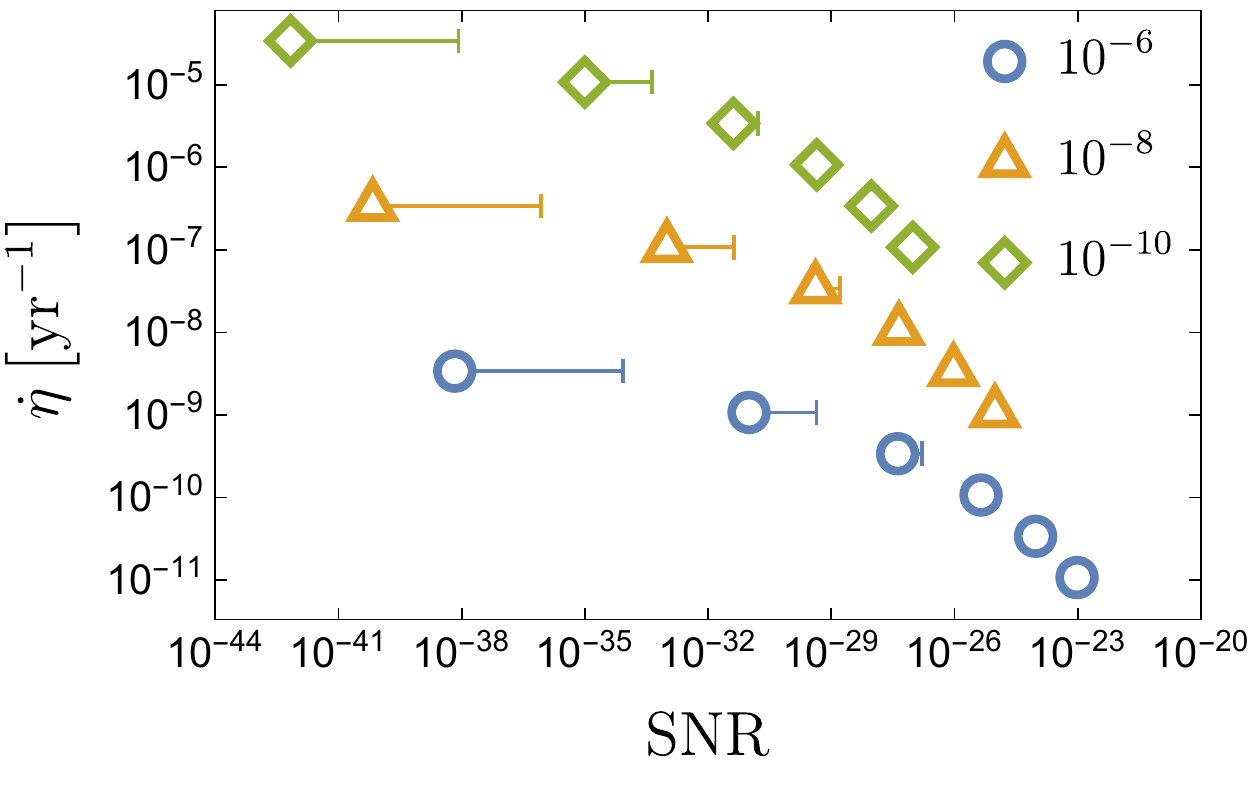}
	\includegraphics[width=0.85\columnwidth]{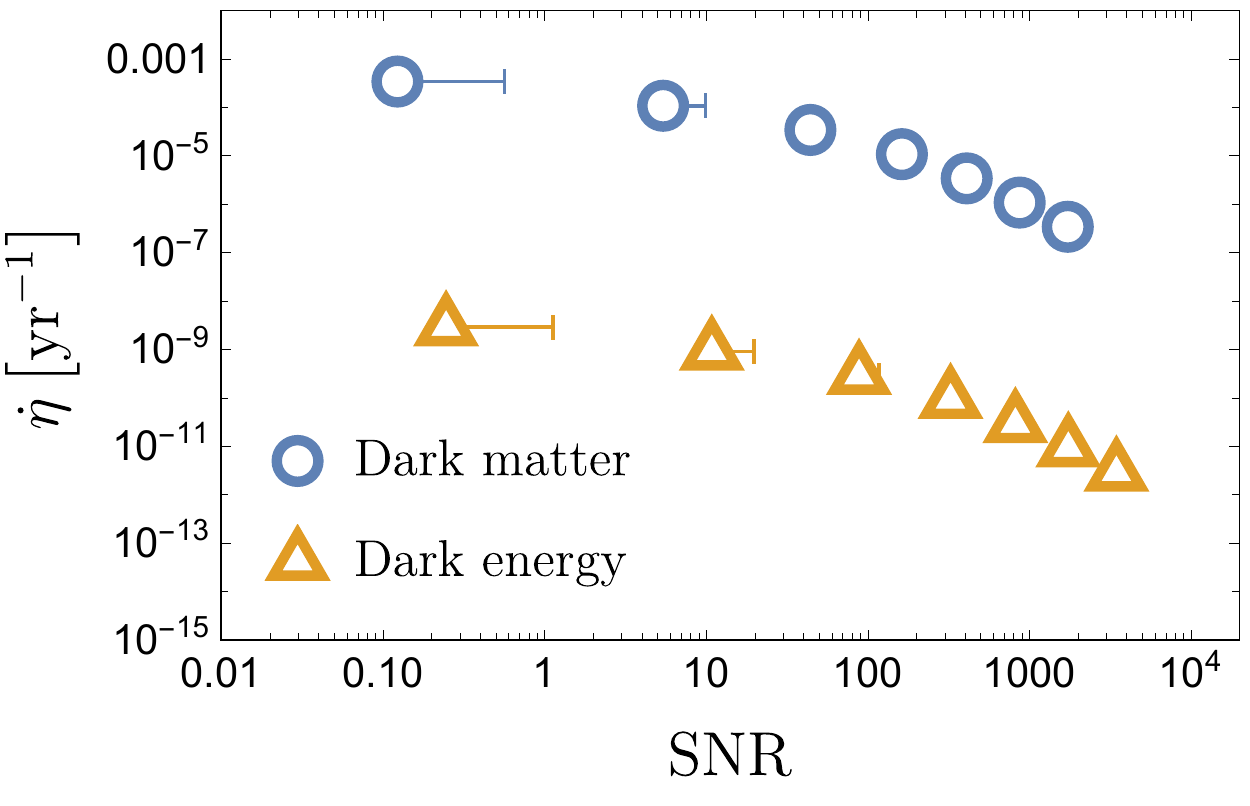}
	\caption{Average gravitational interaction rate of cosmic strings with the interferometer as a function of the signal-to-noise ratio. We distinguish between a network of static cosmic strings (top) and an interacting network of dynamical strings (bottom), that constitutes the dark matter or dark energy component of the universe. In the top panel, the colors denote typical values of the dimensionless string tension, $G\mu/c^2$, while we have fixed the energy density to the dark matter one, $\rho_{\mathrm{S}} \approx 0.39 \, \mathrm{GeV/cm^3}$. In the bottom panel, we have chosen a string tension of $G\mu/c^2 \approx 10^{-23}$ and fixed the energy density and equation of state for dark matter as ($\rho_{\mathrm{DM}} \approx 0.39 \, \mathrm{GeV/cm^3}$, $w=0$) and for dark energy as ($\rho_{\mathrm{DE}} \approx 3.2 \, \mathrm{keV/cm^3}$, $w=-1$). In both panels we have assumed a Maxwell-Boltzmann distribution for the velocity of the strings with root mean square $v_{\mathrm{rms}} \approx 270 \, \mathrm{km/s}$ in combination with the LISA experiment moving through the galaxy at $v_{\odot} \approx 220 \, \mathrm{km/s}$.}
\label{fig:CS-Signal2}
\end{figure}

Given the angular-averaged signal power spectra, let us now estimate the discovery potential for cosmic strings in the vicinity of the gravitational wave interferometer.
Assuming that the energy stored in the cosmic string network is determined by a single scale $d$, that is the characteristic distance between the strings, the energy density is given by $\rho_{\mathrm{S}} \sim \mu / d^2$.
In addition, this scale will also determine the average interaction rate of a string with the detector.
Since we have averaged over the angles describing the orientation of the string, for simplicity, we can assume that the strings are all aligned and moving into the same direction perpendicular to their orientation.
Projecting into the plane orthogonal to the strings then effectively reduces the problem of estimating the flux of cosmic strings to computing the flux of point-particles through a given surface element.
Here, however, we have projected out one dimension and are thus dealing with the flux through a line element.
Accordingly, the rate of cosmic strings approaching the detector at velocity $V$ and impact parameter $D$ can be estimated to be
\begin{equation}
    \dot{\eta} \sim DV \frac{\rho_{\mathrm{S}}}{\mu} \, .
\end{equation}
Note that, in line with the angular-averaged signal power spectral density shown in Fig.~\ref{fig:CS-Signal2}, $\dot{\eta}$ is an estimate of the average rate at which a cosmic string of velocity $V$ is passing by the detector at a distance $D$.
Therefore, it is only an approximation, as it does not take the shape of the signal into account, which, for instance, can crucially depend on the relative orientation between the string and the interferometer (this is similar to the discussion of dark matter clumps in Section~\ref{sec:dmclumps}).
Nevertheless, we do not expect any significant changes of our conclusions, if this complication was taken into account more carefully.

The panels of Fig.~\ref{fig:CS-Signal2} show the average gravitational interaction rate of a cosmic string network of static (top) and interacting strings for $w=0$ and $w=-1$ (bottom; for the yellow curve we also choose the energy density to be that of dark energy, $\rho_{\mathrm{DE}} \approx 3.2 \, \mathrm{keV/cm^3}$~\cite{Aghanim:2018eyx}, which will use throughout this work) with the LISA interferometer as a function of the signal-to-noise ratio.
The bars illustrate the alternative low-frequency cutoff for the noise power spectrum of the detector (cf.~Section~\ref{sec:dmclumps} for details).
For static cosmic strings, we find that, due to their tiny gravitational field, the rate of sufficiently strong interactions with the detector is basically negligible, even for the most optimistic values of the string tension.
Naively, one could of course consider situations where one tries to probe the gravitational field with more massive test objects, thereby increasing the source of the field.
This, however, is completely determined by the experimental setup, such as LISA in our case, and not a free parameter.
Consequently, the tiny gravitational field renders the signal of a static cosmic string by its gravitational pull on the interferometer unobservable.

The more interesting case is an interacting network with an equation of state $w=0$, suitable for being the dominant component of the dark matter.
Due to the greatly enhanced gravitational field, the situation is significantly improved, as we demonstrate in the bottom panel of Fig.~\ref{fig:CS-Signal2}.
For instance, we find that a network with strings of tension $G\mu/c^2 \approx 10^{-23}$ could induce a signal in the interferometer with a signal-to-noise ratio of about 10 every 10,000 years on average.
Taking a more optimistic low-frequency cutoff this could even be improved, as indicated by the bars in the figure.
Therefore, an overall gain in experimental sensitivity could allow to gain a sizeable factor in the signal event rate, giving hope that already moderate improvements will allow for a detection.
As a consequence, similar to the case of dark matter clumps, our analysis strategy would benefit from improvements in the low frequency regime.
Note that, in order to obtain a decent interaction rate with a signal-to-noise ratio still greater than one, we have, to some extent, chosen the string tension close to an optimal value, $G\mu/c^2 \approx 10^{-23}$.
We will show the overall experimental sensitivity of LISA for different string tensions in Fig.~\ref{fig:Conclusion} of Section~\ref{sec:conclusions}.

We conclude that the overall discovery potential of a cosmic string network with the LISA interferometer depends on its dynamics, i.e.~its equation of state.
Our estimates of the signal-to-noise ratio indicate that a network of static vacuum strings appears not to be observable, while an interacting network with an equation of state appropriate for dark matter may be closer to the experimental reach of LISA.
Furthermore, our comparison of different low-frequency cutoffs suggests that the overall discovery potential may be increased by improvements in the low frequency regime.

\subsection{Domain walls}
\label{subsec:DomainWalls}

In order to obtain the gravitational field sourced by a domain wall, we can solve Poisson's equation~\eqref{eq:PoissonNewton} for an energy density confined to an infinite plane, e.g., $\rho = \sigma \delta(x)$.
Here, $\sigma$ denotes the surface tension of the domain wall and we have neglected a possibly finite thickness.
In this background, the gravitational potential grows linearly with the distance, $\phi \sim r$.
Therefore, the gravitational field of a domain wall with an arbitrary orientation is constant,
\begin{equation}
	\vec{g}(\vec{r}) = \mp 2 \pi G (1+3w) \sigma \vec{n} \, .
\label{eq:DWGravitationalField}
\end{equation}
Here, $\vec{n}$ is the unit vector normal to the plane parametrizing the domain wall and the sign ensures that the field always points towards (or, as we will see momentarily, even away from) the wall\footnote{From a geometric point of view, the correct sign for a test mass located at $\vec{r}$ is given by the sign of $\vec{n} \cdot \left( \vec{r} - \vec{r}_0 \right)$, where $\vec{r}_0$ is any point in the plane parametrizing the domain wall.}.
That is, the gravitational field of a domain wall is pointing in the direction normal to it and, in particular, is independent of the distance to the wall.

Similar to cosmic strings, a network of domain walls can have different equations of state, depending on its dynamics.
For instance, the equation of state of a \emph{static} domain wall is given by $w=-2/3$, such that its energy density dilutes as $\rho_{\mathrm{DW}} \sim a^{-1}$, while a domain wall moving at a velocity $\beta = v/c$ obeys an equation of state of $w = \beta^2 - 2/3$~\cite{Kolb:1990vq}.
Intriguingly, according to~\eqref{eq:DWGravitationalField}, this implies that the gravitational field of a static domain wall is repulsive rather than attractive~\cite{Vilenkin:1981zs,Vilenkin:1984hy}.
Nevertheless, complicated dynamics of an interacting network of domain walls may lead to a very different equation of state.
Therefore, it can as well serve as an exotic candidate for dark matter or dark energy~\cite{Bucher:1998mh,Battye:1999eq,Friedland:2002qs} with a corresponding equation of state.

In the following, we want to derive the signal due to the gravitational pull (or push, in the case of a repulsive potential) of a domain wall travelling through the LISA interferometer.

\subsubsection*{Signal power spectrum}

The gravitational field of a domain wall is independent of the distance to it.
In contrast to the detection of dark matter clumps or cosmic strings, this means that all three nodes of the interferometer will experience the same acceleration due to the presence of a domain wall.
Therefore, in order to obtain a differential acceleration and hence a signal in the interferometer, the domain wall has to traverse the space between the different satellites.
That is, it has to completely separate one spacecraft from the other two, thereby accelerating them into opposite directions.

The velocity perturbation that each node picks up due to the constant acceleration in the gravitational field of the domain wall reads
\begin{equation}
	\vec{v} (t) = \mp 2\pi G (1+3w) \sigma t \Theta\left(t\right) \vec{n} \, ,
\label{eq:DW_vi}
\end{equation}
where $\vec{n}$ denotes the normal vector of the domain wall and the different signs follow the conventions of~\eqref{eq:DWGravitationalField}.
Here, for simplicity, we assume that the wall starts traversing the interferometer at a time $t_0=0$.
Hence, the $\Theta$-function implements the fact that, due to the equal acceleration of the three nodes, we do not expect a signal, if the domain wall does not separate the individual nodes from each other.
Similarly, we will assume that the signal ceases to exist when the domain wall has traversed the detector volume completely.
In other words, for simplicity, we consider a signal induced by a domain wall, that starts travelling through the detector by passing the first node, then passing the second, which thereafter gets accelerated in the opposite direction, and finally traverses the third node, after which it ceases to exist.
That is, strictly speaking, we view the signal as caused by the acceleration burst instead of the individual velocity perturbations.
When considering the latter, obviously, the signal will not cease to exist after the domain wall has passed the last satellite, as there is still a differential velocity shift between all three nodes.
In principle, this will, in addition, lead to a persistent deformation of the LISA triangle.
In general, the situation is even more complicated, when considering a complex network of interacting domain walls which traverse the detector volume at different times and directions.
A correct and thorough treatment would require an involved numerical simulation of this scenario.
Here, we will not consider this layer of complexity.

Due to the constant gravitational field, the signal that a domain wall will induce in the interferometer when \linebreak traversing the detector volume involves each of its three nodes.
Therefore, the detector response has to be \linebreak parametrized by the exact response function given in~\eqref{eq:ResponseFunctionFullAppendix}, where the $\vec{v}_i$ are now given by~\eqref{eq:DW_vi}.
The terms of the detector response function can be evaluated explicitly, e.g., in a reference frame where each node is located on a coordinate axis, $\vec{r}_i = L/\sqrt{2} \vec{e}_i$, and we parametrize the domain wall by the unit vector $\vec{n} = \left( \sin\theta \cos\phi, \sin\theta \sin\phi, \cos\theta \right)$.
We can then proceed by considering the Fourier transform of the response function, i.e.~of the velocity perturbations, and uniformly averaging over the angles $(\theta, \phi)$ in order to obtain the angular-averaged signal power spectrum.
For a detailed discussion of the geometrical aspects of this, see Appendix~\ref{app:GeomPlanes}.
We also note that, similar to our discussion of localized dark matter clumps, a uniform average is only an approximation.
We discuss its validity and more details in Appendix~\ref{app:dmvelocity}.

\begin{figure}[t]
\centering
	\includegraphics[width=0.85\columnwidth]{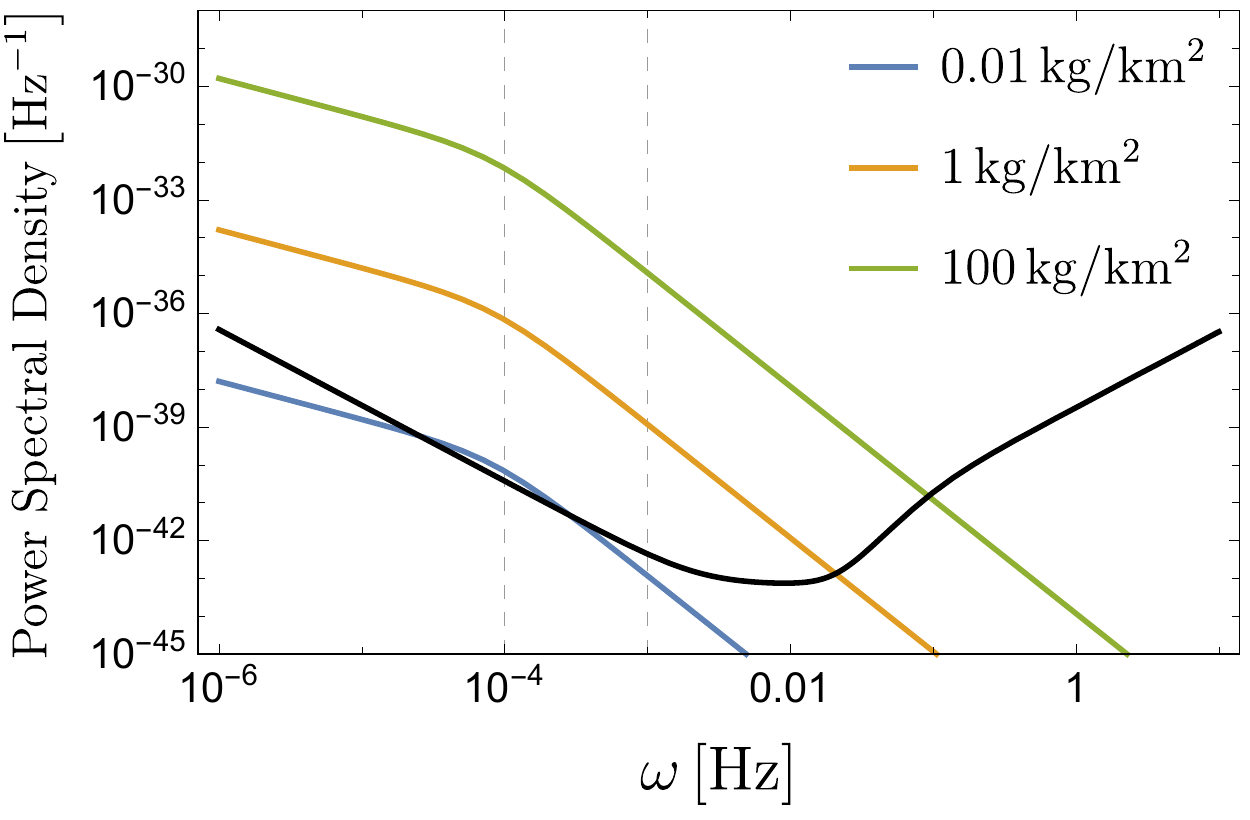}
	\caption{Angular-averaged signal power spectral density of a domain wall traversing the detector volume compared to the experimental sensitivity of LISA. The signal is shown in color, illustrating different surface tensions of the domain wall, $\sigma$. The sensitivity of LISA is shown in black. Again the thin dashed line indicate two choices of the low-frequency cutoff. In contrast to the case of clumps and strings, the lower one is chosen at a smaller frequency as our calculation for the domain walls has no close-approach limitation in the present case. Here, we assume that the domain wall is moving at a velocity of $v/c = 10^{-3}$. If the domain wall network is to cover the dark matter component of the Universe, the signal events shown here are quite infrequent. On average, we expect them every 1.5 (blue), 150 (yellow) and, in the extreme case, 15,000 (green) years.}
\label{fig:DW-Signal}
\end{figure}

In Fig.~\ref{fig:DW-Signal} we illustrate the angular-averaged signal power spectral density due to the differential gravitational acceleration by a domain wall traversing the interferometer and compare it to the experimental sensitivity of LISA.
In particular, we consider domain walls of different surface tensions and fix their typical velocity to $v/c = 10^{-3}$.
We find that, compared to the case of dark matter clumps or cosmic strings (cf.~Fig.~\ref{fig:DM-Signal} and Fig.~\ref{fig:CS-Signal2}), the signal caused by a domain wall is significantly enhanced in the high frequency regime of LISA's characteristic frequency range.
One factor contributing to this is that we assumed an infinitely thin domain wall and point like nodes of the experiment.
In practice, both have a finite thickness that should lead to a faster drop off at large frequencies.
However, this does not play any significant role in our determination of the sensitivity.

Furthermore, we can estimate the average rate of gravitational interactions with the LISA detector that we expect for a given network of domain walls.
The energy density that is stored in a network is essentially determined by the surface tension of the walls $\sigma$ and the characteristic distance $d$ between them, $\rho_{\mathrm{DW}} \sim \sigma / d$.
In order to estimate the rate at which we expect them to approach the detector, we can consider the simplified situation where the domain walls of the network are all parallel to each other\footnote{We recall that, in our simple approximation, we have already uniformly averaged over their direction of motion (cf.~Appendix~\ref{app:GeomPlanes}).}.
The overall rate is then simply given by the inverse time interval between two consecutive walls travelling through the detector volume.
That is, if the domain walls move at a certain velocity $v$, the rate at which we expect them to approach the detector can be roughly estimated by
\begin{equation}
    \dot{\eta} \sim v \frac{\rho_{\mathrm{DW}}}{\sigma} \, .
\end{equation}
Note that, in line with the angular-averaged signal power spectral density shown in Fig.~\ref{fig:DW-Signal}, $\dot{\eta}$ is merely an estimate of the average rate at which a domain wall is passing through the detector and is therefore only an approximation.
In particular, it does not take the shape of the signal caused by a specific domain wall into account, which, for instance, can crucially depend on the relative orientation between the domain wall and the detector (this is similar to the our discussion of dark matter clumps in Section~\ref{sec:dmclumps}).
As an extreme example, a domain wall that is exactly parallel to the interferometer plane would not leave any experimental imprint in the detector.
While this does not make any significant difference for the results we show in this section, we present results of using a somewhat improved approach in Fig.~\ref{fig:Conclusion} of Section~\ref{sec:conclusions}.
There we account for this additional angular dependence.
In particular, a thorough treatment requires to weight the signal events according to their relative orientation with respect to the detector.
For instance, as a simple schematic example, this can be written (see also Appendix~\ref{app:GeomPlanes})
\begin{equation}
    \dot{\eta}_{\mathrm{tot}} = \frac{1}{4\pi} \int_{0}^{\pi} \diff \theta \sin\theta \int_{0}^{2\pi} \diff \phi \, \dot{\eta} \, \Theta \left[ \mathrm{SNR} (\theta, \phi) - s \right] \, ,
\label{eq:DWRateExact}
\end{equation}
where $s$ is the required signal-to-noise ratio.
That is, naively, only signal events are taken into account for configurations which lead to a detectable signal at the interferometer.
Nevertheless, this does not qualitatively alter the results presented in this section.

\begin{figure}[t]
\centering
	\includegraphics[width=0.85\columnwidth]{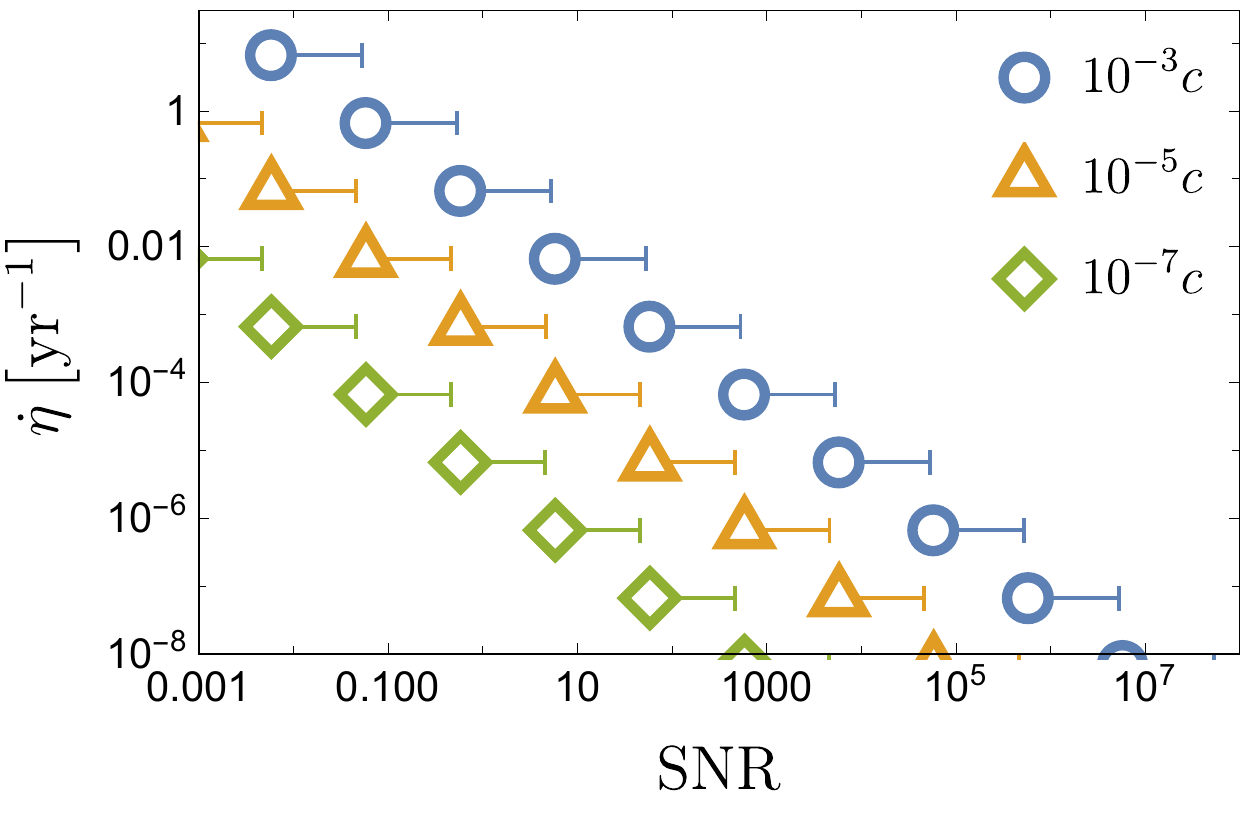}
	\includegraphics[width=0.85\columnwidth]{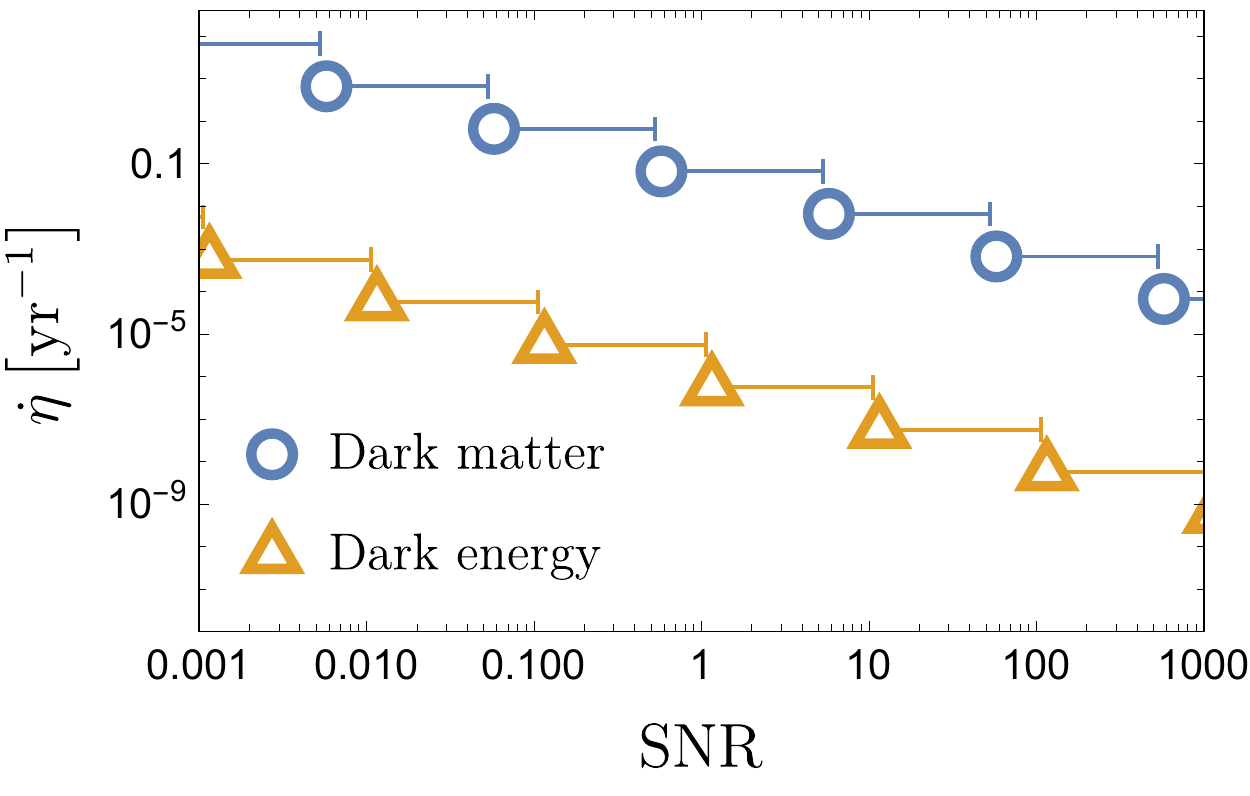}
	\caption{Average gravitational interaction rate of a static domain wall network (top) and an interacting domain wall network (bottom) as a function of the signal-to-noise ratio. In the top panel, the colors denote the velocity at which the wall is traversing the detector. Here, we choose the domain wall energy density to be of the same order of magnitude as dark matter. In the bottom panel, we assume a typical velocity of the domain walls of $v/c=10^{-3}$ together with an energy density and equation of state corresponding to dark matter (blue, $\rho_{\mathrm{DW}} \approx 0.39 \, \mathrm{GeV/cm^3}$, $w=0$) and dark energy (yellow, $\rho_{\mathrm{DW}} \approx 3.2 \, \mathrm{keV/cm^3}$, $w=-1$), respectively. Note that, for a given interaction rate, the domain wall tension is fixed by the combination of the velocity and the dark matter density.}
\label{fig:DW-SNR}
\end{figure}

Fig.~\ref{fig:DW-SNR} shows the average gravitational interaction rate of a domain wall network as a function of the signal-to-noise ratio.
In the top panel, we consider the case of a static domain wall network, while in the bottom panel, we consider an interacting network of domain walls that constitute the dark matter or dark energy component of the Universe.
Similar to Fig.~\ref{fig:DM-Eta}, the bars illustrate an alternative choice of the low-frequency cutoff for the noise power spectrum of the detector given in~\eqref{eq:NoisePowerSpectrum}.
However, as our estimate of the signal induced by a domain wall does not rely on the close-approach approximation, the cutoff can be shifted towards even lower frequencies.
Here, we choose $\omega_c = 10^{-4} \, \mathrm{Hz}$ as indicated in Fig.~\ref{fig:DW-Signal}.
For both setups, we observe that with increasing signal-to-noise ratio the event rate reduces according to a linear power law.
This is because, here, we consider the angular-averaged signal-to-noise ratio, $\mathrm{SNR} \propto \sigma$, as well as interaction rate, $\dot{\eta} \propto \sigma^{-1}$, which are solely determined by the domain wall surface tension $\sigma$.
That is, each data point shown in the figure corresponds to a specific value of the latter.

In general, as expected, the overall discovery potential for static and interacting domain walls is comparable, as long as the energy density is fixed to that of dark matter.
Intriguingly, it is also somewhat better than in the case of dark matter clumps or cosmic strings.
For instance, we expect a signal due to a domain wall traversing the interferometer with a signal-to-noise ratio of about ten every 10 to 100 years on average.
Taking into account potential improvements due to a more optimistic noise spectrum, this rate might also be higher.
In this case the same signal may even be expected almost every 1 to 10 years on average.
Therefore, in the future, such a domain wall network could certainly be within experimental reach of LISA.

Let us close this discussion with a few words of caution.
The estimates presented in this section are subject to some simplifications we have made in our derivation.
Our results are based on the assumption that the domain wall can be parametrized by an infinite plane of vanishing thickness.
This assumption, however, might not always be fully justified in particle physics models that admit domain wall solutions.
Moreover, and perhaps more importantly, we have only considered the situation of a domain wall inducing a signal that ceases to exist once the wall has traversed the entire detector volume.
In principle, one would also have to include the remnant velocities of the nodes relative to each other once the domain wall has passed, which most likely would modify the signal power spectrum at low frequencies.
In addition, this would also change the triangular detector geometry persistently.
Nevertheless, our relatively conservative estimates for gravitational interactions of domain wall networks with the LISA interferometer provide hope for a future discovery potential.

\section{Stochastic Fluctuations of the Dark Matter Density}
\label{sec:dmfluctuations}

In addition to strongly localized energy densities, such as dark matter clumps or topological defects travelling through the Universe, the gravitational potential in the vicinity of the interferometer can also be perturbed by stochastic fluctuations of the dark matter density, as already discussed in~\cite{Baghram:2011is} for the example of pulsar timing arrays.
To linear order, these gravitational perturbations satisfy
\begin{equation}
	\Laplace \delta\phi = 4 \pi G \bar{\rho} \delta \, ,
\label{eq:PoissonFluctuations}
\end{equation}
where $\delta \phi$ denotes the perturbation of the gravitational potential, while $\delta = \delta \rho / \bar{\rho}$ is the relative fluctuation of the mean density $\bar{\rho}$.

In principle, (time dependent) perturbations of the gravitational potential are precisely what gravitational wave experiments try to measure.
In this section, we aim to estimate the sensitivity of LISA to these perturbations caused by fluctuations of the dark matter density.
For simplicity, we consider a simple density fluctuation, oscillating in space, that travels through the interferometer.
That is, after decomposing $\delta$ in Fourier space, the gravitational response to a Fourier mode with wave vector $\vec{k}$ is given by
\begin{equation}
	\tilde{\delta\phi}_{\vec{k}} = - 4 \pi G \bar{\rho} \frac{\tilde{\delta}_{\vec{k}}}{k^2} \, ,
\end{equation}
where $k = \abs{\vec{k}}$.

\begin{figure}[t]
\centering
	\includegraphics[width=0.85\columnwidth]{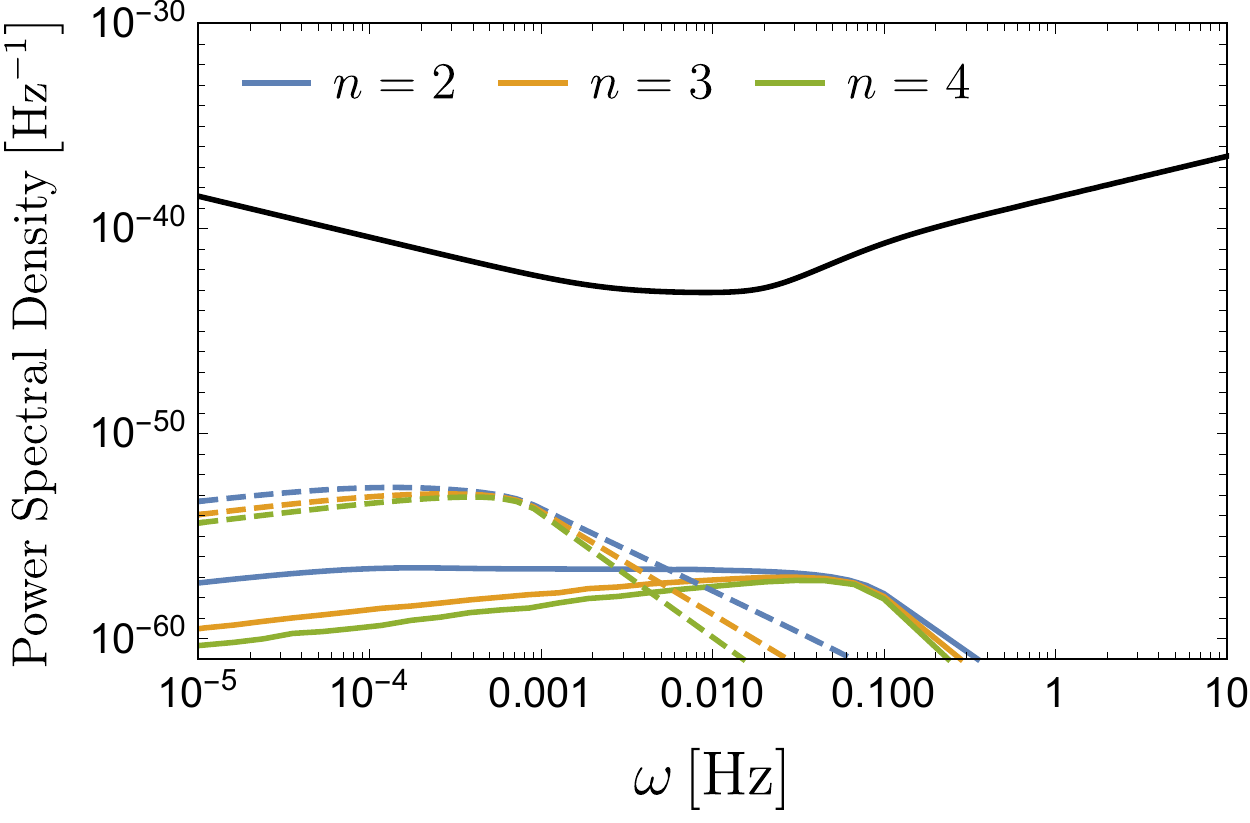}
	\caption{Angular-averaged power spectral density of a local dark matter density fluctuation compared to the experimental sensitivity of LISA. The colors denote different spectral indices $n$. Amongst these, the solid and dashed lines correspond to the peak positions, $k_{\star}$, of the dark matter power spectrum, i.e.~$k_{\star}^{-1} \sim 2.2 \times 10^{3} \, \mathrm{km}$ (solid) and $k_{\star}^{-1} \sim 2.2 \times 10^{5} \, \mathrm{km}$ (dashed), respectively. The black line shows the sensitivity of LISA. Here, we have fixed the normalization of the dark matter power spectrum to $\mathcal{P}_{\star}=1$. We furthermore assume a velocity of $v\approx 220 \, \mathrm{km/s}$ at which the detector is moving through the density wave, while for the dark matter density we use the same value as in Fig.~\ref{fig:DM-Signal}.}
\label{fig:Stochastic-Signal}
\end{figure}

We can now determine the detector response to a specific Fourier mode of the density fluctuation by considering the contribution to the gravitational perturbation of each mode separately, $\delta \vec{g}(\vec{x}) = \int \diff^3 k \, \delta \vec{g}_{\vec{k}} (\vec{x})$, where the perturbation of the gravitational field associated to a mode with wave vector $\vec{k}$ reads
\begin{equation}
	\delta \vec{g}_{\vec{k}} (\vec{x}) = 4 \pi i G \bar{\rho} \tilde{\delta}_{\vec{k}} \frac{\vec{k}}{k^2} \exp \left( i \vec{k} \cdot \vec{x} \right) \, .
\end{equation}
Naively, this corresponds to a situation, where we select a plane wave of specific wave length and ``freeze it", as there is no dynamical wave evolution.
Instead, we obtain a detector response to this density wave, if the interferometer moves through the perturbation of the gravitational field.
To some extent, this setup is similar to a network of domain walls travelling through the interferometer, discussed in Section~\ref{subsec:DomainWalls}.

Assuming a detector node is moving through the density perturbation of wave vector $\vec{k}$ at a constant velocity $\vec{v}$, i.e.~$\vec{x} (t) = \vec{v}t + \vec{x}_0$, the frequency spectrum of the associated gravitational acceleration is given by
\begin{equation}
	\tilde{\delta \vec{g}}_{\vec{k}} (\omega) = 4 \pi i G \bar{\rho} \tilde{\delta}_{\vec{k}} \frac{\vec{k}}{k^2} \sqrt{2\pi} \exp \left( i \vec{k} \cdot \vec{x}_0 \right) \delta \left( \omega + \vec{k} \cdot \vec{v} \right) \, .
\end{equation}
Note that the right most term denotes the $\delta$-distribution, not to be confused with the density fluctuation.
This represents that, if the detector moves through a plane wave at a constant velocity, there is only one frequency contributing, which depends on the angle between the wave vector of the density perturbation and the motion of the detector.

The above gravitational acceleration will, in turn, lead to a velocity perturbation of each node.
Similar to the discussion of domain walls in Section~\ref{subsec:DomainWalls}, we can plug each velocity perturbation into the detector response function, $X(t)$, to obtain the response of the interferometer to a given Fourier mode of a density perturbation, $\tilde{X}_{\vec{k}} (\omega)$.
Furthermore, in order to determine the detector response to a superposition of density fluctuations, we can then sum over all Fourier modes.
The absolute square of this finally yields the signal power spectrum associated to the linear superposition of density perturbations, which schematically reads
\begin{equation}
\begin{split}
	P(\omega) &= \int \diff\omega^{\prime} \diff^3 k \, \diff^3 k^{\prime} \, \avg{\tilde{X}_{\vec{k}^{\prime}}^{\ast} (\omega^{\prime}) \tilde{X}_{\vec{k}} (\omega)} \\
	& \sim 32 \pi^3 \frac{G^2 \bar{\rho}^2}{c^2} \sum_{i,j=1}^3 c_{ij}(\omega) \\
	& \times \int \diff\omega^{\prime} \diff^3 k \, \diff^3 k^{\prime} \, \frac{\left( \vec{n}_i \cdot \vec{k}\right) \left(\vec{n}_j \cdot \vec{k}^{\prime}\right)}{\omega \omega^{\prime} k^2 k^{\prime 2}} \\
	& \times \me^{i \vec{k} \cdot \vec{r}_i} \me^{-i \vec{k}^{\prime} \cdot \vec{r}_j} \delta \left( \omega + \vec{k} \cdot \vec{v} \right) \delta \left( \omega^{\prime} + \vec{k}^{\prime} \cdot \vec{v} \right) \avg{\tilde{\delta}^{\ast}_{\vec{k}^{\prime}} \tilde{\delta}_{\vec{k}}} \, ,
\end{split}
\end{equation}
where, similar to the previous sections, $\vec{n}_i$ denote the unit vectors pointing between two nodes, labelled by the opposite side of the triangle, and $\vec{r}_i$ is the initial position of each node.
The coefficients $c_{ij}(\omega)$ encode the linear combination of the velocity perturbations of each interferometer node in the detector response~\eqref{eq:ResponseFunctionFullAppendix}.
Therefore, they also include phases of the form $\exp(i\omega L/c)$ due to retardation effects.

That is, naively, the signal response of the detector to a superposition of density fluctuations with different wavelengths is weighted according to the dark matter density power spectrum.
For the latter we use the conventional definition (see, e.g.,~\cite{Dodelson:2003ft})
\begin{equation}
	\avg{\tilde{\delta}^{\ast}_{\vec{k}^{\prime}} \tilde{\delta}_{\vec{k}}} = \left( 2 \pi \right)^3 \delta^{(3)}( \vec{k} - \vec{k}^{\prime}) \frac{2\pi^2}{k^3} \mathcal{P}(k) \, .
\end{equation}

As a simple example of the density power spectrum, we consider a broken power law\footnote{In particle physics models of very light dark matter, such a spectrum and the wavelengths of interest to us may be achieved quite naturally, see, e.g.,~\cite{Graham:2015rva,Cosme:2018nly,Alonso-Alvarez:2018tus}.},
\begin{equation}
	\mathcal{P}(k) = \mathcal{P}_{\star} \left( \Theta (k_{\star} - k) \left( \frac{k}{k_{\star}} \right)^n + \Theta (k - k_{\star}) \left( \frac{k}{k_{\star}} \right)^{-n}\right) \, ,
\end{equation}
where $\mathcal{P}_{\star}$ is a normalization constant and $n$ is the spectral index.
Obviously, $\mathcal{P}(k)$ exhibits a peak $\mathcal{P}_{\star}$ at a characteristic scale $k_{\star}$, which we treat as a free parameter.
Indeed, for the purpose of this work, we assume that it obtains its maximum within the characteristic frequency band of LISA,  $\omega_{\star}$.
The corresponding wavelength is then of the order
\begin{equation}
	k_{\star}^{-1} \sim 2.2 \times 10^{5} \, \mathrm{km} \left( \frac{v}{220 \, \mathrm{km/s}} \right) \left( \frac{1 \, \mathrm{mHz}}{\omega_{\star}} \right) \, .
\end{equation}
Again we average over angles (cf.~Appendices~\ref{app:GeomPlanes} and~\ref{app:dmvelocity}).
We illustrate the angular-averaged signal power spectral density associated to a dark matter density fluctuation traversing the interferometer in Fig.~\ref{fig:Stochastic-Signal} and compare it to the overall experimental sensitivity of LISA.
The signal power spectrum shown is normalized to unity, i.e.~$\mathcal{P}_{\star}=1$.
We also show different choices of the spectral index $n$ as well as different peak positions $k_{\star}$ of the dark matter power spectrum.
For simplicity, we have fixed the velocity at which the detector is moving through the density wave to $v \approx 220 \, \mathrm{km/s}$.
For density fluctuations of order 1 we find that, even for optimistic choices of the peaks, the overall signal power spectrum is suppressed by many orders of magnitude compared to the noise.
As the signal scales linearly with the power spectrum of the fluctuations, $\mathcal{P}_{\star}$, we can see that enormous fluctuations are needed for a signal to be detectable.
Indeed, this result is, to some extent, expected from our discussion of localized dark matter clumps.
Naively, clumps correspond to a fluctuation of significant overdensity compared to the background.
However, as we have shown in Section~\ref{sec:dmclumps}, these are scarcely within experimental reach of the LISA detector.
In summary, we therefore do not expect any detectable experimental signature of stochastic fluctuations of the dark matter density at LISA, unless the overdensities are very large (at which point they may actually resemble more the localized structures we have already discussed).

\section{Extrapolation to Other Gravitational Wave Experiments}
\label{sec:GWInterferometers}

The general strategy we have presented in the previous sections is not limited to the LISA interferometer, but in principle applies to any experimental apparatus that is sensitive to gravitational perturbations.
In this section, we want to extrapolate our results to other gravitational wave experiments.
In particular, we aim to obtain the detectable rate corresponding to the gravitational pulls of dark matter clumps and topological defects in ground-based interferometers, for example LIGO~\cite{Abbott:2007kv,TheLIGOScientific:2014jea}, and pulsar timing arrays, for instance utilizing SKA~\cite{Smits:2008cf}.
Due to their different characteristic sizes compared to LISA, these detectors are sensitive to gravitational interactions in other frequency regimes and therefore, ultimately, to other energy densities of dark matter objects.

\subsection{LIGO}
\label{subsec:LIGO}

With its pioneering measurements, LIGO has strongly advanced the field of gravitational wave astronomy~\cite{Abramovici:1992ah,Abbott:2016blz}.
For the purpose of this work, we will treat it as a classical Michelson interferometer, which aims to measure phase shifts between laser beams sent across two orthogonal arms.
Since the characteristic length of both arms is $L \approx 4 \, \mathrm{km}$, it is sensitive to gravitational perturbations at frequencies of approximately $1 \, \mathrm{Hz}$ to $1000 \, \mathrm{Hz}$~\cite{Abbott:2007kv,TheLIGOScientific:2014jea}.

In general, while the basic idea of our survey for LISA also applies to LIGO, there are a few differences due to the detector geometry.
Most importantly, LIGO is a ground-based interferometer.
Naively, its detector nodes are freely hanging mirrors suspended from a static laboratory frame, in stark contrast to the freely moving satellites of LISA.
Therefore, while the detector response at LISA was determined by relative velocity shifts between the satellites, at LIGO we can only expect a signal due to sufficiently short gravitational acceleration bursts on the mirrors.
Consequently, as a simplified detector response to a gravitational perturbation, we consider the differential acceleration between the four mirrors~\cite{Saulson:1995zi},
\begin{equation}
	X(t) = \left[ g_x^1(t) - g_x^0(t) \right] - \left[ g_y^1(t) - g_y^0(t) \right] \, ,
\end{equation}
where $x$ and $y$ denote the arms of the interferometer.
In addition, being a ground-based interferometer, LIGO is subject to different sources of background noise, such as seismic motion, for example.
To take the different background noise into account, we model the noise power spectrum according to the sensitivity curve of the design sensitivity of advanced LIGO~\cite{TheLIGOScientific:2014jea}.
Since we parametrize the detector response $X(t)$ in terms of differential accelerations, we then convert the strain- to an acceleration noise by multiplying with $(\omega^2 L / 2)^2$ (see, e.g.,~\cite{Maggiore:1900zz}).
With these definitions, we can repeat the analysis of the previous sections.

\begin{figure}[t]
\centering
	\subfloat[Dark matter clumps]{
		\centering
		\includegraphics[width=0.78\columnwidth]{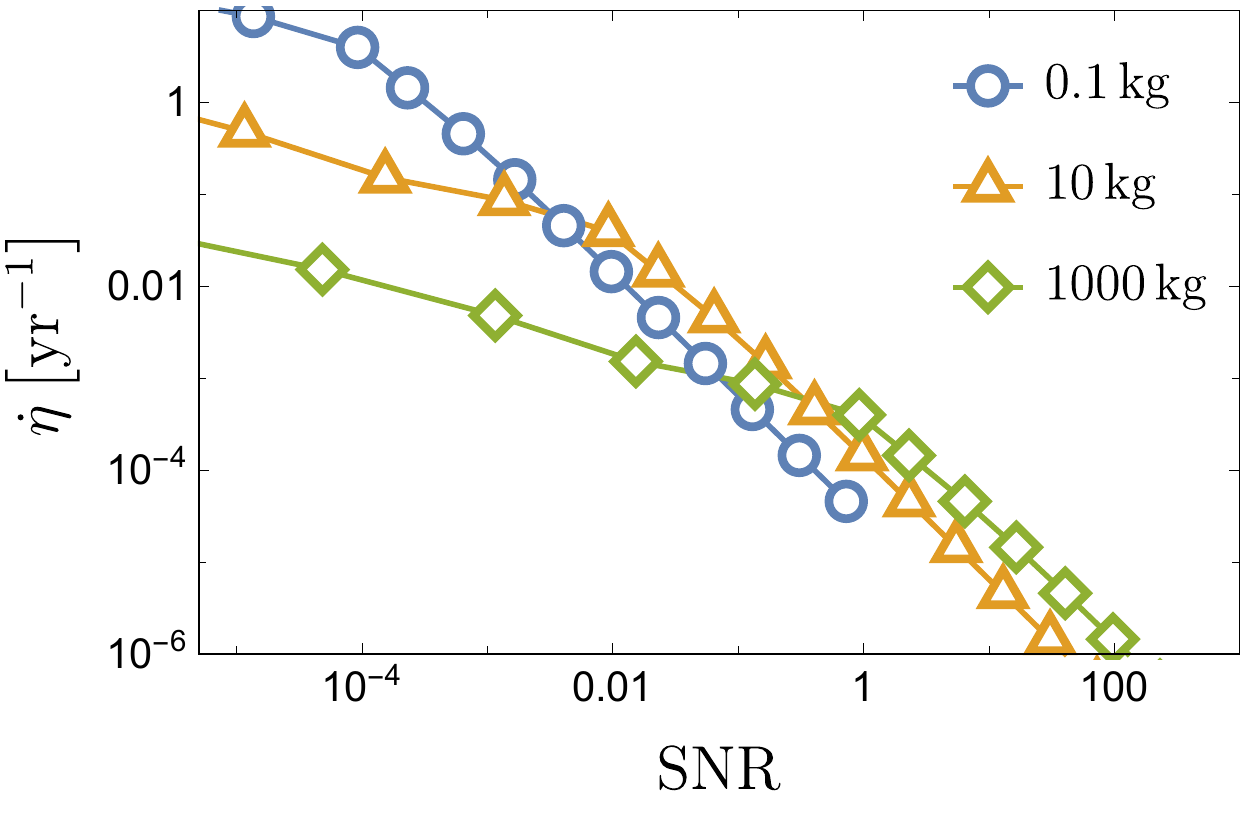}
	}
	\\
	\subfloat[Cosmic strings]{
		\centering
		\includegraphics[width=0.78\columnwidth]{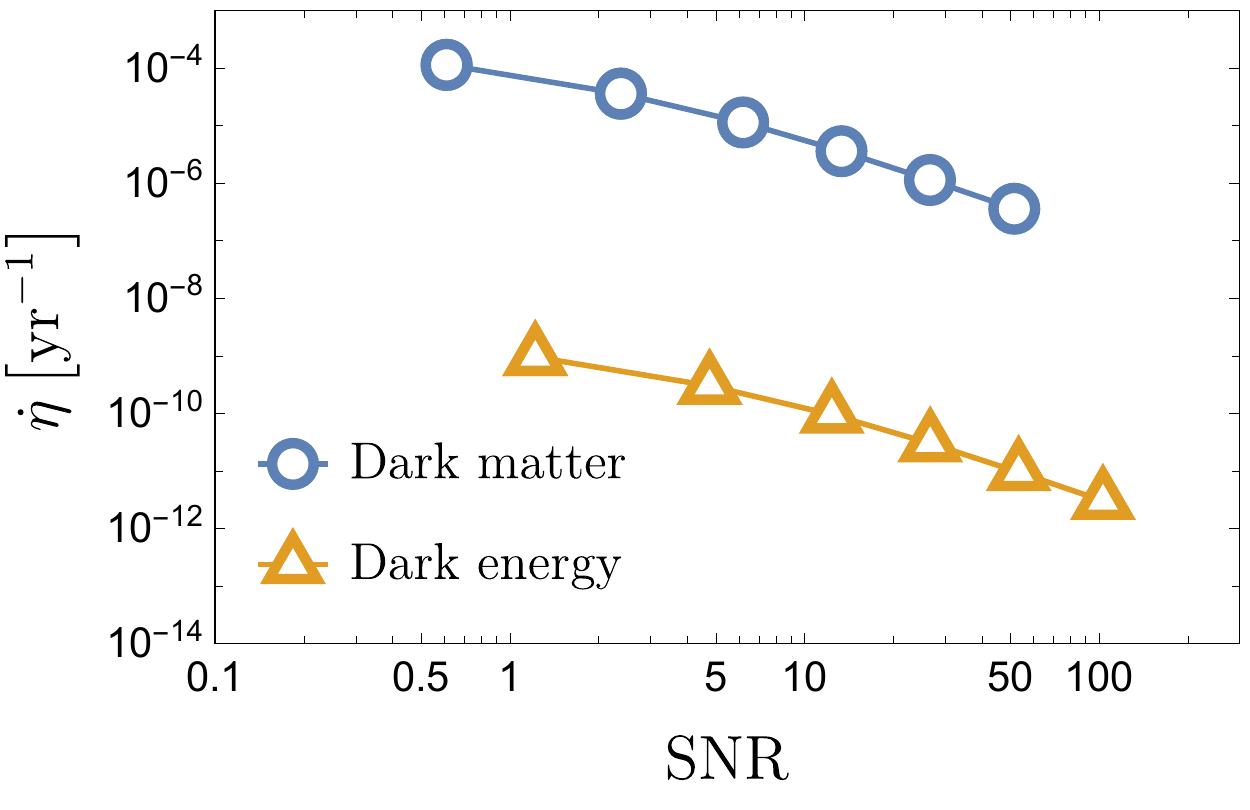}
	}
	\\
	\subfloat[Domain walls]{
		\centering
		\includegraphics[width=0.78\columnwidth]{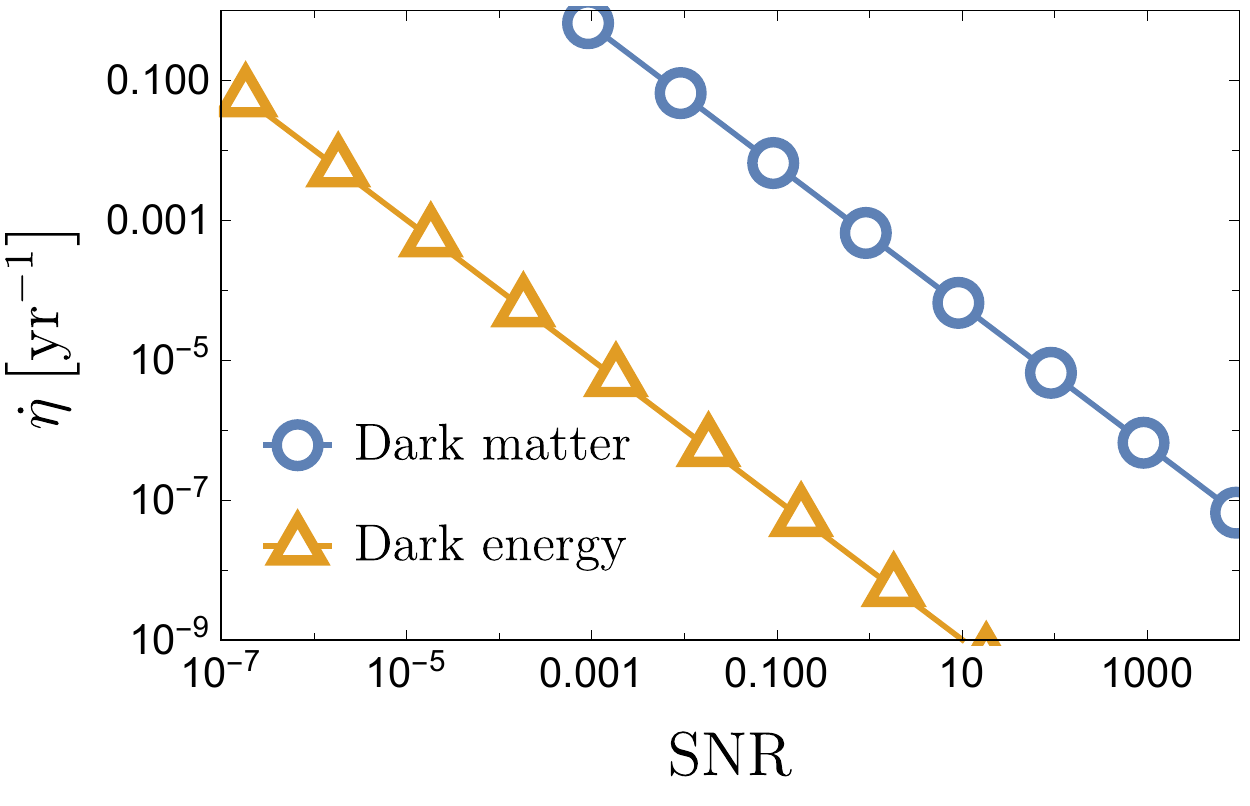}
	}
\caption{Average gravitational interaction rate of localized dark matter clumps (a), cosmic strings (b) and domain walls (c) with LIGO as a function of the signal-to-noise ratio. (a) The colors illustrate different masses of the clumps. (b) Here the color indicates the equation of state of the cosmic string network. We furthermore assume strings of tension $G\mu/c^2 \approx 10^{-28}$. (c) As in (b) but for a domain wall network travelling at a fixed velocity, $v/c = 10^{-3}$ (as noted in Fig.~\ref{fig:DW-SNR}, the combination of interaction rate, density and velocity fixes the domain wall tension). All remaining parameters for dark matter and dark energy, as well as for the velocity distribution of clumps and strings, are chosen as in Fig.~\ref{fig:CS-Signal2}.}
\label{fig:SNRvEtaLIGO}
\end{figure}

We summarize our results in Fig.~\ref{fig:SNRvEtaLIGO}, where we show the average gravitational interaction rate of localized dark matter clumps as well as string and domain wall networks with LIGO as a function of the signal-to-noise ratio.
Note that here, due to the small characteristic size of the detector as compared to LISA, we cannot use the close-approach approximation of the detector response, $D \lesssim L$, in every region of the parameter space we are interested in.
For simplicity, and in order to also take the regime $D \gg L$ into account, we extrapolate our results by multiplying the gravitational acceleration by a factor $L/D$ in this limit (see~\eqref{eq:ResponseRegimes} for a qualitative explanation).

In general, for localized dark matter clumps, we observe a behaviour qualitatively similar to the LISA analysis shown in Fig.~\ref{fig:DM-Eta}.
As one may expect, we find that LIGO with its much smaller size is sensitive to clumps with comparatively low masses.
We find that dark matter clumps of masses $10^{3} \, \mathrm{kg}$, could be observed every 2,000 years on average.
More optimistically, given that these estimates crucially depend on the noise spectrum as well as the frequency cut-offs used to determine the signal-to-noise ratio (in addition to the other approximations we employed), these values could also be a factor of 10 or even higher.
Bearing this in mind, our results are in rough agreement with~\cite{Hall:2016usm}.
An overview of the expected signal event rate for different masses of the clumps is shown in Fig.~\ref{fig:Conclusion}.

We obtain similar results for networks of topological defects.
As an example, we show a cosmic string network with tension $G\mu/c^2 \approx 10^{-28}$, for which we find that LIGO could observe a corresponding signal with a signal-to-noise ratio above one approximately every 20,000 years on average.
Note that these values depend on the string tension and velocity distribution we assume.
The signals induced by a domain wall network traversing the interferometer are similar to the ones we obtained at LISA.
Quantitatively, we expect a possible signal with a signal-to-noise ratio above one every 1,000 to 10,000 years on average.

In summary, LIGO can measure sufficiently short gravitational acceleration bursts caused by localized clumps of dark matter or networks of cosmic strings and domain walls.
Due to LIGO's smaller size and correspondingly different frequency range compared to LISA, it is sensitive to objects of typically smaller masses as well as string and domain wall tensions.
Indeed, with the same analysis strategy, further experimental improvements are required to lift the discovery potential of LIGO for these dark matter structures to a level that allows for a detection in a reasonable time frame.

\subsection{Pulsar timing arrays}
\label{subsec:PTAs}

Another type of experiment, aiming towards a measurement of a gravitational waves, are pulsar timing arrays (PTAs).
These exploit the fact that the time of arrival of radiation bursts from pulsars across the Universe can be precisely predicted.
Gravitational waves that traverse the space between a certain set of pulsars and the observer on Earth will perturb the correlations between the different times of arrival, thereby generating a signal in the corresponding detector.
Due to the long lines of sight between the pulsars and the telescopes, PTAs are able to probe gravitational wave spectra in very low frequency regimes, typically of the order $10^{-9} \, \mathrm{Hz}$ to $10^{-6} \, \mathrm{Hz}$ (see, e.g.,~\cite{FosterPTA,Smits:2008cf,Jenet:2009hk,Hobbs:2009yy,Ferdman:2010xq,Manchester:2012za}).

In this section we want to explore the discovery potential of PTAs with respect to the astrophysical structures discussed in the previous sections, i.e.~dark matter clumps (note the previous works~\cite{Siegel:2007fz,Seto:2007kj,Baghram:2011is,Kashiyama:2012qz,Clark:2015sha,Schutz:2016khr,Kashiyama:2018gsh,Dror:2019twh}) as well as networks of cosmic strings and domain walls.
For simplicity, we will do so by viewing a PTA as a gravitational wave interferometer of the same type as common ground- or space-based interferometers at very large scales, i.e.~as an interferometer with arm lengths of several thousand light-years.
As a prototypical example, we will consider a replica of LISA, that is an interferometer made up by three nodes at equal distance to each other, but with an arm length of $L \approx 1000 \, \mathrm{ly}$.
However, the sensitivity of PTAs is usually based on the observation of multiple pulsars. To use this, the signal must exist in all Earth-pulsar combinations. For our search strategy this requires that the dark matter structure is affecting the velocity of the Earth. Compared to LISA, where all nodes can be treated equally, this reduces the detectable rate by a factor of 3.
Although this is a crude simplification of the experimental techniques used for gravitational wave measurements with PTAs, we still expect reasonable estimates of prospective signal strengths for the purpose of this work.
In particular, in terms of average signal event rates, we expect to benefit from the largely increased detector volume.

As we are essentially considering a LISA experiment at large scales, our analysis strategy is similar to the one presented in Sections~\ref{sec:dmclumps} and~\ref{sec:topologicalobjects}.
The only major difference is the sensitivity of the experiments in different frequency regimes.
We implement this by modelling the noise power spectrum according to the typical sensitivity curve of a PTA.
As a prototype, we use the noise power spectrum of a PTA utilizing the future Square Kilometer Array (SKA)~\cite{Smits:2008cf} as shown in~\cite{Moore:2014lga},
\begin{equation}
\begin{split}
	S_n^{\mathrm{SKA}} (f) \approx & \frac{1}{2\pi} 8.6 \times 10^{-18} \left(\frac{f}{1\,\mathrm{Hz}}\right) \mathrm{Hz^{-1}} \\
	& \times \left( \frac{2 \left(2\pi f L / c\right)^2}{1 + \left(2 \pi f L / c\right)^2} \right)^2 \, , \quad f \geq f_c \, ,
\end{split}
\end{equation}
and extrapolate it to high frequencies.
Here, similar to~\eqref{eq:NoisePowerSpectrum}, we again introduced a transfer function to adapt the raw strain spectrum to our signal spectrum.
Furthermore, we added an additional factor of $(2\pi)^{-1}$ in order to match the Fourier conventions.
Due to the limited observation time of SKA, we introduce a cut-off for frequencies below $f_c \approx 1.5 \times 10^{-9} \, \mathrm{Hz}$ in the computation of the corresponding signal-to-noise ratio.

\begin{figure}[t]
\centering
	\subfloat[Dark matter clumps]{
		\centering
		\includegraphics[width=0.78\columnwidth]{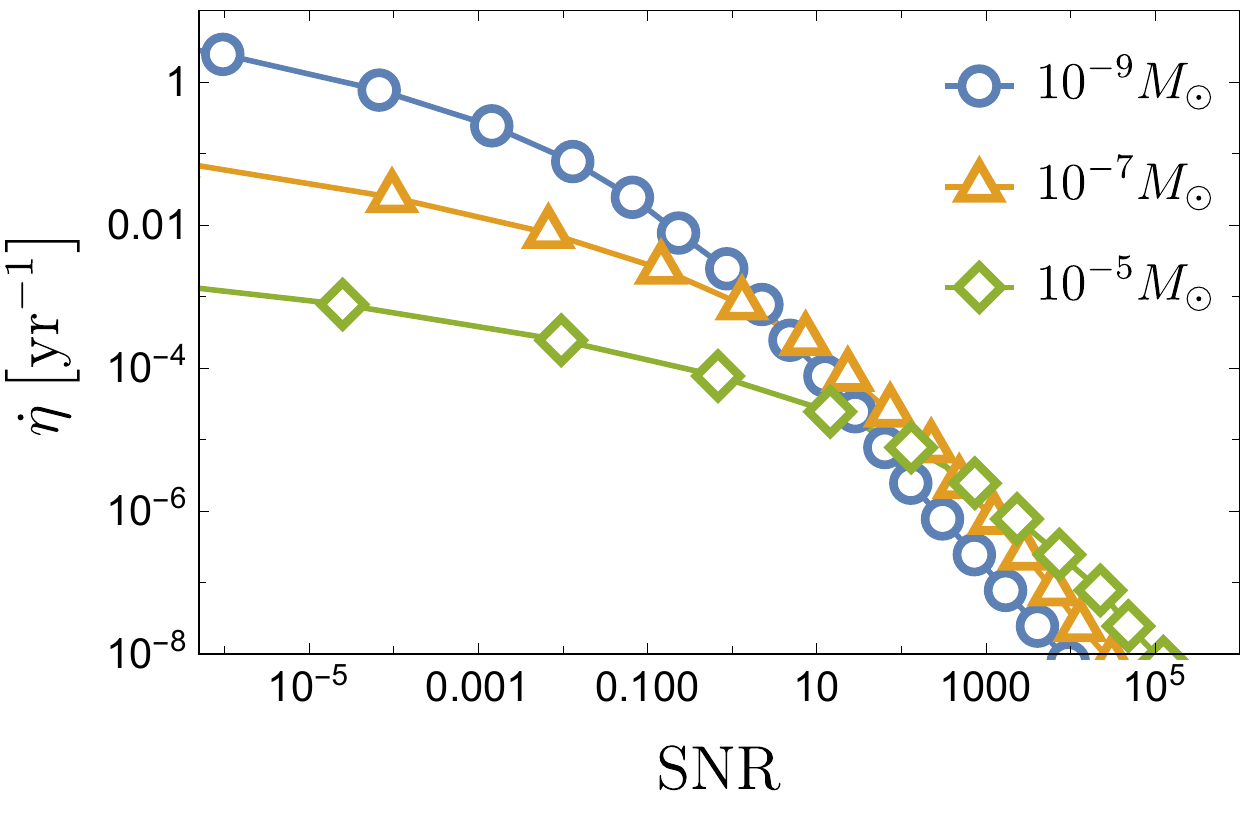}
	}
	\\
	\subfloat[Cosmic strings]{
		\centering
		\includegraphics[width=0.78\columnwidth]{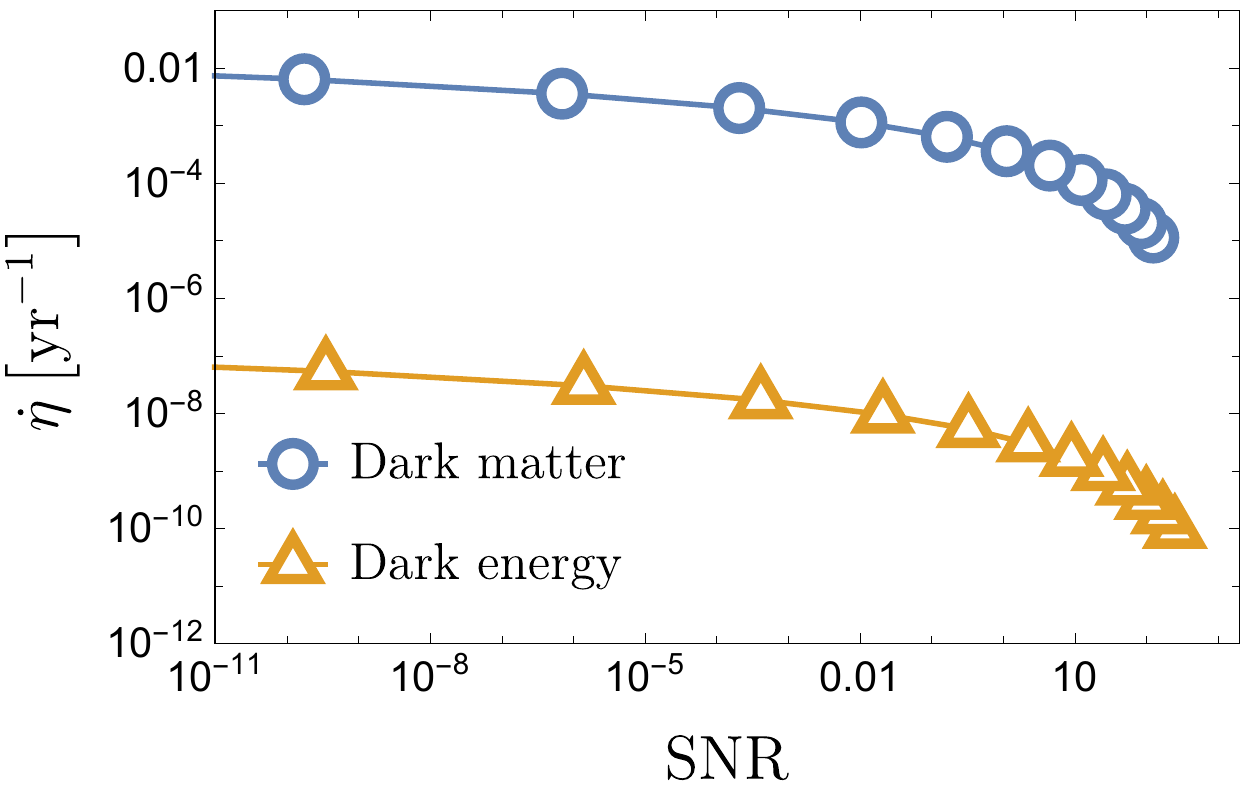}
	}
	\\
	\subfloat[Domain walls]{
		\centering
		\includegraphics[width=0.78\columnwidth]{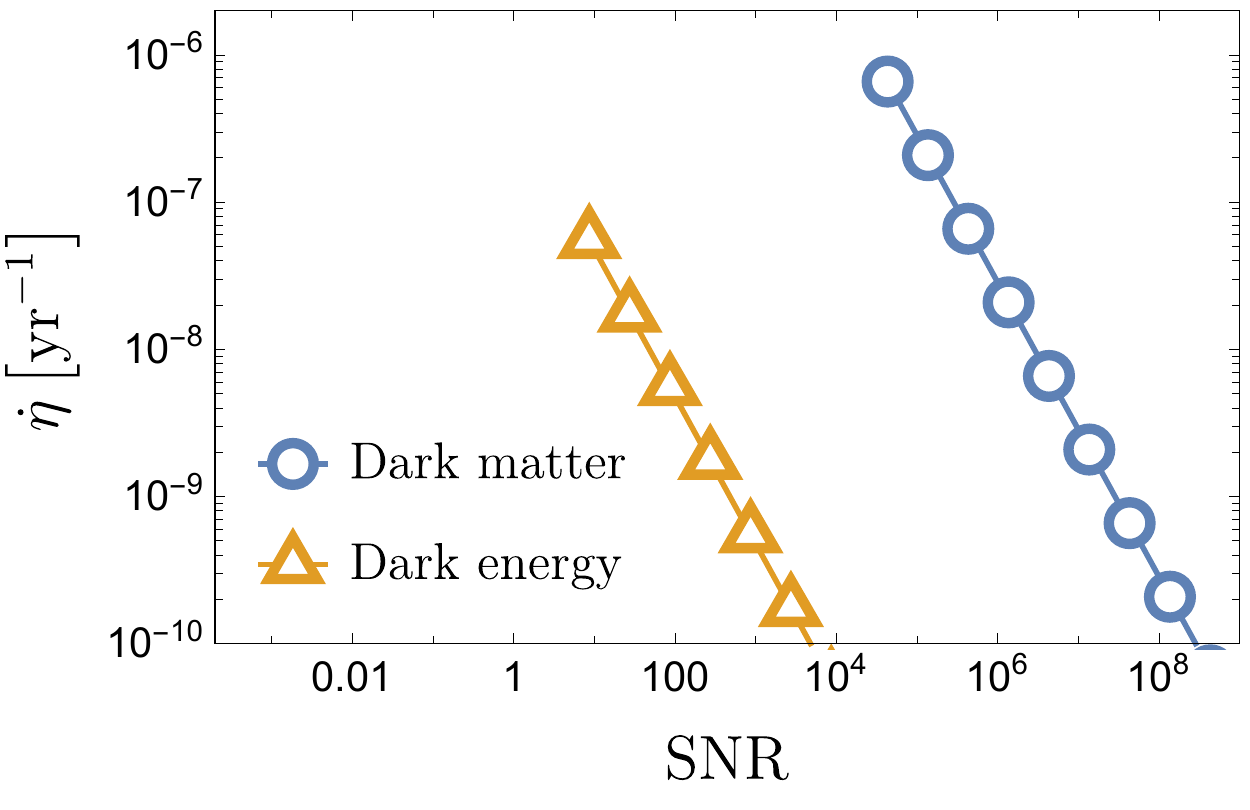}
	}
\caption{As Fig.~\ref{fig:SNRvEtaLIGO} but for a prototypical PTA utilizing SKA. All parameters are also the same except for the string tension in (b), which is chosen to be $G \mu / c^2 \approx 10^{-18}$.}
\label{fig:SNRvEtaSKA}
\end{figure}

A summary of our results is given in Fig.~\ref{fig:SNRvEtaSKA}, where we show the average gravitational interaction rate of localized dark matter clumps as well as string and domain wall networks as a function of the signal-to-noise ratio at SKA.
Note that here, similar to the analysis of LISA, we consider the close-approach regime of the localized dark matter clumps and cosmic strings.
That is, we only consider events with impact parameters smaller than the size of the detector, $D \lesssim L$.
Because of the large size of the experiment, this condition is basically fulfilled in any region of the parameter space we are interested in.

For localized clumps of dark matter, we find that we benefit from the increased detector volume, allowing for very massive clumps while still retaining decent interaction rates.
As expected, the qualitative behaviour we observe is similar to the detection of dark matter clumps with LISA shown in Fig.~\ref{fig:DM-Eta}.
Due to the increased detector volume, we expect a discovery potential for signals induced dark matter clumps with masses of $M_{\mathrm{DM}} \approx 10^{-9} M_{\odot}$ at a signal-to-noise ratio of $\mathrm{SNR} \sim 1$ every 300 years on average.
Apparently, this is a slightly more pessimistic estimate compared to what was found in earlier works.
For example, in~\cite{Seto:2007kj} the original estimate of a detection rate for primordial black hole dark matter with PTAs is somewhat higher.
Nevertheless, bearing in mind the crude approximations we have applied in our derivation, our results are in rough agreement.

We find a similar discovery potential for networks of cosmic strings as well as domain walls, constituting the dark matter or dark energy budget of the Universe, respectively.
In particular, a cosmic string network featuring a larger string tension appears to benefit from the experimental setup at large scales.
Here, we show an example of strings with a tension of $G \mu / c^2 \approx 10^{-18}$.
For this network we expect a signal with a signal-to-noise ratio of about 1-10 every 1000 to 10,000 years on average.
For a domain wall network with the same equation of state, we find an expected signal with a signal-to-noise ratio of $10^5$ every few million years on average.
We clearly do not need such a high signal-to-noise ratio.
However, to preserve the validity of our analysis we require to have only a single domain wall traversing the detector volume at the same time.
This, in turn, implies a lower bound on the surface tension of the domain wall network.
As each data point shown is uniquely determined by a specific surface tension, this forbids us to go beyond the shown signal-to-noise ratios and corresponding interaction rates.
Nevertheless, keeping these caveats in mind, we can extrapolate the given data.
This raises the hope that we can have events every 10 to 100 years with a signal-to-noise ratio above one, $\mathrm{SNR} \gtrsim 1$.

In summary, we find that, in a scenario where the dark matter properties optimally fit the experimental sensitivity, dark matter clumps, cosmic strings and domain walls, may be observed at SKA with a slightly higher rate than LIGO while being comparable to LISA.
All categories benefit from the increased detector volume, such that PTAs are sensitive to more massive objects than smaller gravitational wave interferometers.
We will give an overview of this feature in Section~\ref{sec:conclusions}, which then also makes explicit the complementarity that exists between these different types of experiments.
However, carefully note that the values we present in this section are all based on the assumption that a measurement of gravitational perturbations with a PTA is the same as with the LISA detector, but at large scales.
The discovery potential of PTAs for strings and domain walls using a more accurate treatment of its experimental techniques certainly merits further investigation.

\section{Conclusions}
\label{sec:conclusions}

In this paper, we have studied the potential of gravitational wave interferometers to measure gravitational perturbations caused by the presence of macroscopic dark matter objects in the vicinity of the detector.
Any localized energy density passing by sufficiently close to the interferometer will exert a gravitational pull on each of its nodes, and hence cause a differential acceleration.
This acceleration leads to Doppler shift signal in the interferometer that can be measured.

The objects we have considered in the present work include localized clumps of dark matter (see also~\cite{Adams:2004pk,Seto:2004zu,Siegel:2007fz,Seto:2007kj,Berezinsky:2010kq,Baghram:2011is,Kashiyama:2012qz,Berezinsky:2014wya,Clark:2015sha,Hall:2016usm,Kawasaki:2018xak,Schutz:2016khr,Kashiyama:2018gsh,Dror:2019twh,Grote:2019uvn}), topological defects, such as cosmic strings and domain walls, as well as stochastic fluctuations (cf.~\cite{Baghram:2011is}) of the dark matter density.
As our baseline example, we have examined the LISA experiment for which we have given the signal power spectrum associated to the presence to each of these sources in the vicinity of the detector.
Based on this we then looked\footnote{Note that our analysis strategy for LIGO and PTAs corresponds essentially to an extrapolation of LISA to different scales. Therefore, these estimates need to be taken with a bit of caution.} at LIGO and a pulsar timing array using SKA that are sensitive to complementary frequency ranges.

\begin{figure*}[t!]
\centering
	\subfloat[Dark matter clumps]{
		\centering
		\includegraphics[width=0.85\columnwidth]{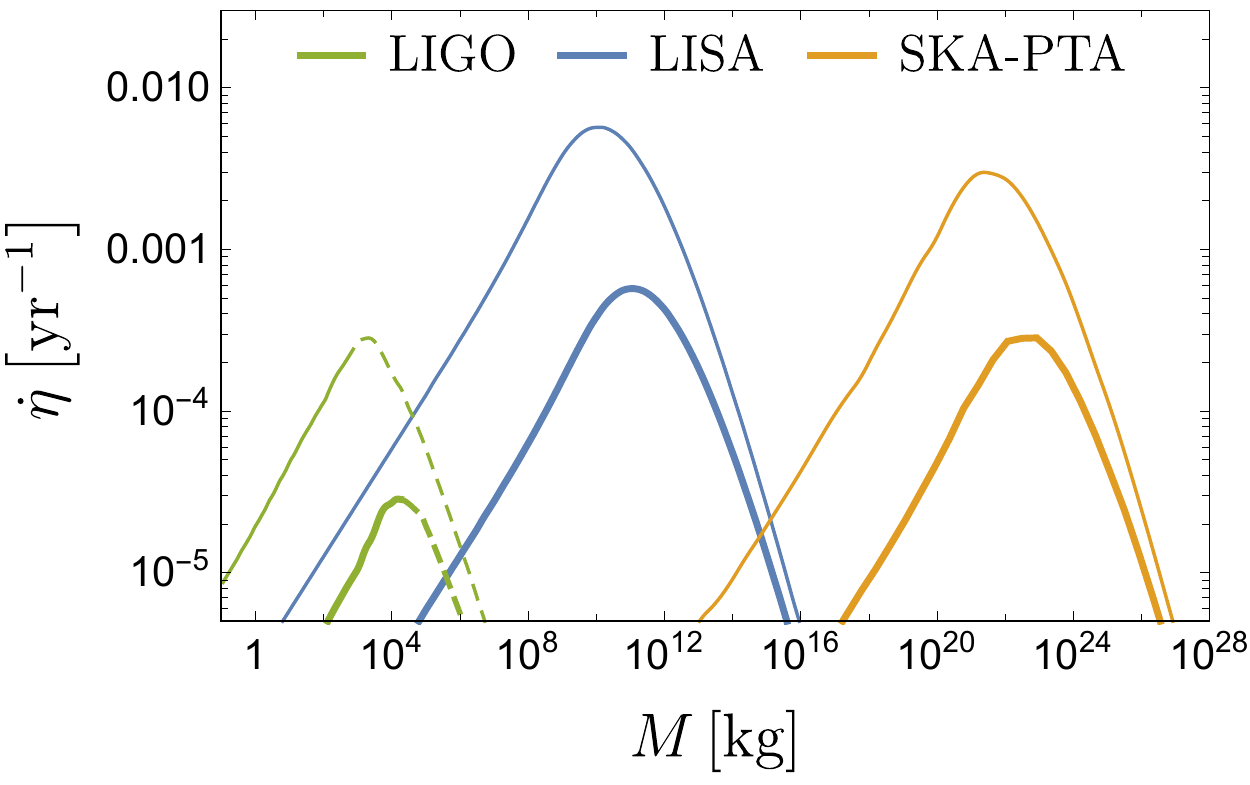}
	}
	\subfloat[Cosmic strings]{
		\centering
		\includegraphics[width=0.85\columnwidth]{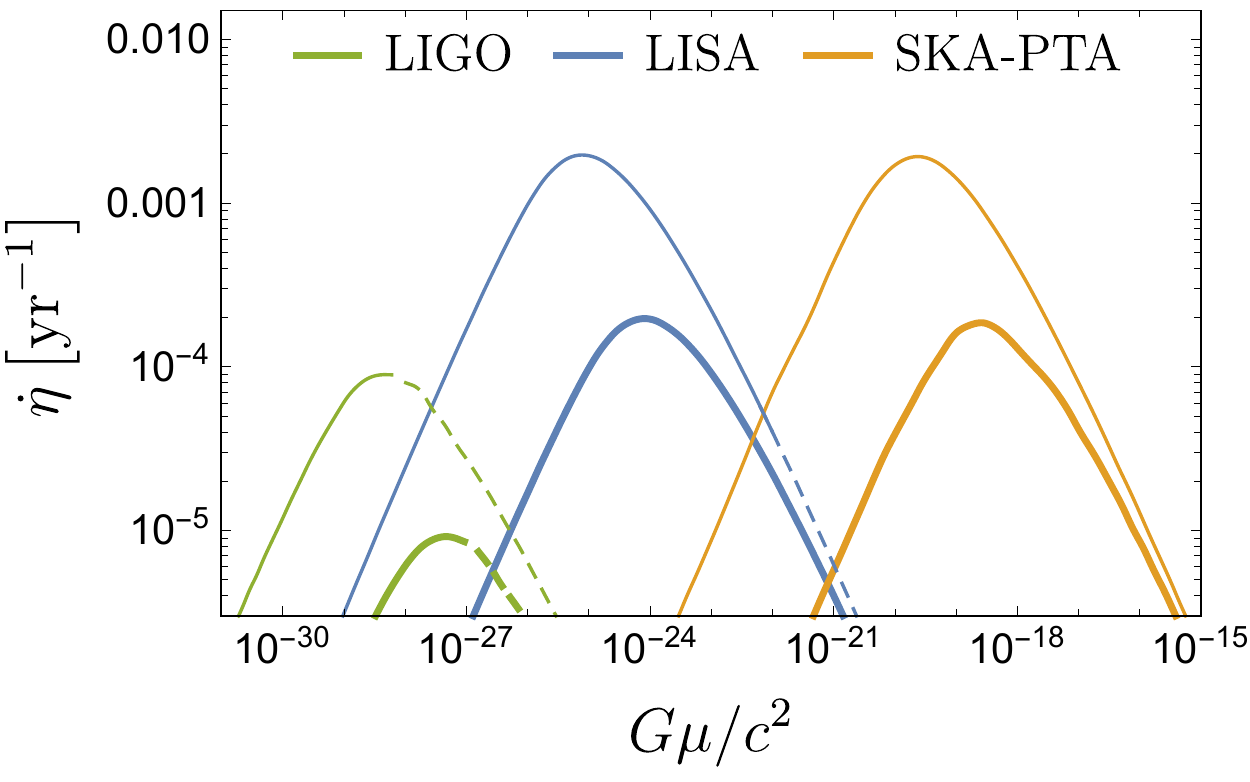}
	}
	\\
	\subfloat[Domain walls]{
		\centering
		\includegraphics[width=0.85\columnwidth]{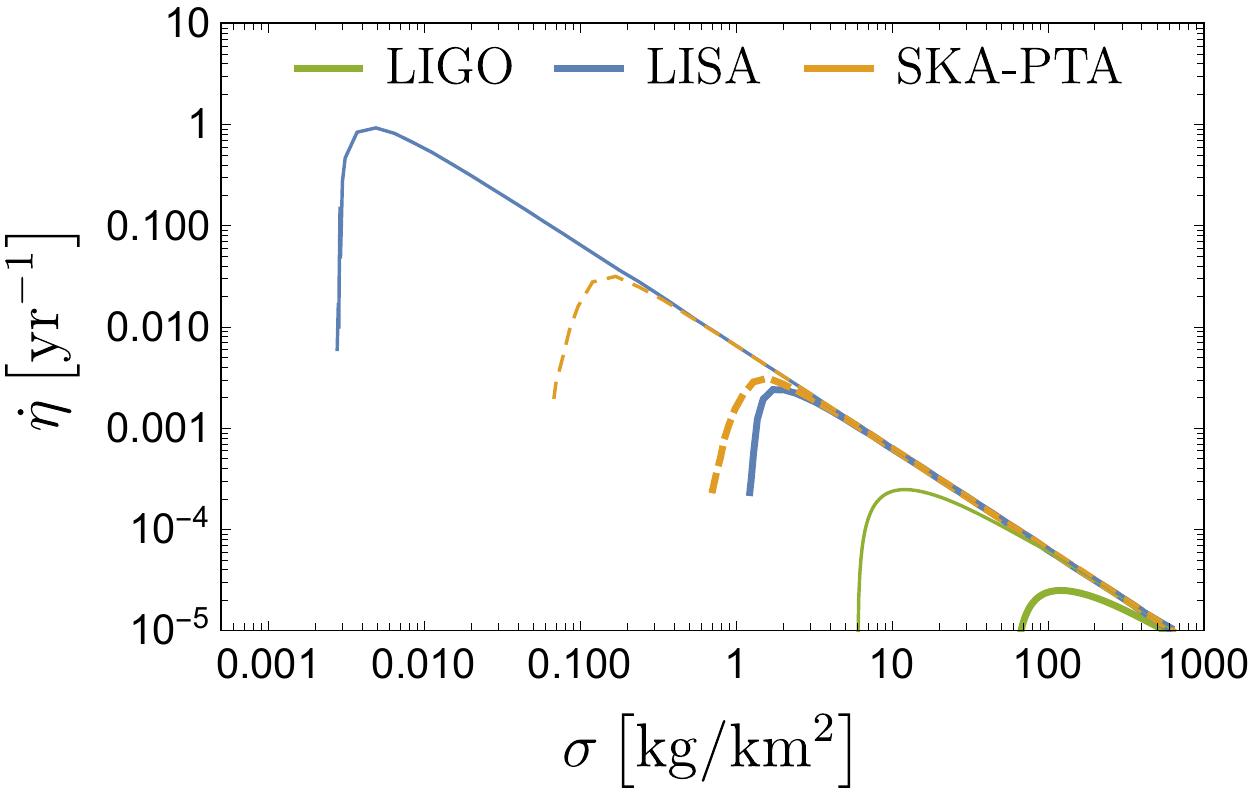}
	}
\caption{Sensitivity of gravitational wave experiments to gravitational perturbations caused by localized dark matter clumps (a), cosmic strings (b) and domain walls (c). The colors illustrate examples of gravitational wave detectors, LIGO, LISA and a prototypical PTA using SKA. The thin lines illustrate a signal-to-noise ratio threshold of $\mathrm{SNR} \gtrsim 1$ while the thicker lines illustrate a case where $\mathrm{SNR} \gtrsim 10$. (a) As in previous figures, we fix $\rho_{\mathrm{DM}} \approx 0.39 \, \mathrm{GeV/cm^3}$ and assume a velocity distribution of Maxwell-Boltzmann type for the dark matter clumps with root mean square $v_{\mathrm{rms}} \approx 270 \, \mathrm{km/s}$ in combination with the experiment moving through the galaxy at $v_{\odot} \approx 220 \, \mathrm{km/s}$. (b) The string network is chosen to constitute the dark matter component of the Universe ($\rho_{\mathrm{DM}} \approx 0.39 \, \mathrm{GeV/cm^3}$, $w=0$). We furthermore assume the strings have the same velocity distribution as the dark matter clumps. (c) Similar to (b) but for a domain wall network at a velocity of $v/c = 10^{-3}$. However, in contrast to (a) and (b) where we used a uniform average, here, the determination of the total rate takes into account the relative orientation between the domain walls and the experiment (see~\eqref{eq:DWRateExact}).
In all panels, a dashed line illustrates a naive extrapolation of our analysis. In (a) and (b) we extrapolate outside the close-approach regime by rescaling the differential acceleration with an additional tidal factor $L/D$ (see~\eqref{eq:ResponseRegimes}). In (c) we extrapolate the signal to a regime, where, on average, there is more than one domain wall traversing the detector volume at the same time.
}
\label{fig:Conclusion}
\end{figure*}

Our results are summarized in Fig.~\ref{fig:Conclusion}.
For each experiment, we show contours of the average gravitational interaction rate of events with a signal-to-noise ratio greater than one, $\mathrm{SNR} \gtrsim 1$, as a function of the energy density of each dark matter structure.
Due to their different characteristic sizes, LIGO, LISA and SKA complement each other very well in the sense that they are sensitive to different masses and tensions of dark matter clumps and topological defects, respectively.

In the most sensitive regimes of each experiment, we find a prospective signal with a signal-to-noise ratio $\mathrm{SNR} \gtrsim 1$, every 10,000 up to 100 years on average.
Note that the most striking signature is a domain wall traversing LISA which might even be expected almost annually.
However, we also note that a signal-to-noise ratio $\mathrm{SNR} \gtrsim 1$ is in fact, at best, a minimum requirement for a signal to be detected.
As the latter still needs to be distinguished from various other sources, a significantly higher signal-to-noise threshold is probably more realistic.
As an example, we illustrate a detection threshold of $\mathrm{SNR} \gtrsim 10$ by the dash-dotted lines in Fig.~\ref{fig:Conclusion}.
While this is not yet on a desirable level for a near future discovery potential, let us remark that the signal estimates we show here are based on the close-approach approximation, imposing a relatively high low-frequency cutoff.
Indeed, crucially, the signal-to-noise ratio can vary significantly, if a different low-frequency cutoff is taken into account (cf.~Section~\ref{sec:dmclumps} for a detailed discussion).
This indicates that improvements of the experimental sensitivity as well as the theoretical analysis, notably in the low-frequency region,
may lead to a sizable enhancement of the detection rate.
Keeping this in mind, localized clumps of dark matter as well as cosmic strings and domain walls may still be within experimental reach of LIGO, LISA and PTAs.
In contrast, as clumps of dark matter clumps naively constitute a critical overdensity of dark matter, stochastic fluctuations of the latter most likely cannot be measured above background noise.
Overall, this clearly requires more sensitive future gravitational wave experiments.

In addition to the acceleration burst signals we have studied in this work, a signal could also arise due to the Shapiro effect~\cite{Shapiro:1964uw}, i.e.~from the changing gravitational potential due to a dark matter structure within the line of sight connecting different nodes of a gravitational wave detector (see~\cite{Siegel:2007fz,Baghram:2011is,Clark:2015sha,Schutz:2016khr,Dror:2019twh}).
It would be interesting to investigate the corresponding signal associated to the presence of cosmic strings or domain walls, which we leave for future work.

\bigskip

In summary, not only localized clumps of dark matter but also cosmic strings or domain walls are close to the experimental reach of gravitational wave interferometers.
Current and future gravitational wave experiments, such as LIGO, LISA and PTAs, are sensitive to gravitational perturbations due to the presence of these objects in the vicinity of the detector.
These experiments are complementary to each other, as the different characteristic sizes and time-scales of the detectors make them sensitive to different parameter regions of the gravitational sources.
Already moderate improvements in the detector noise and analysis may yield interesting discovery potential to intriguingly exotic dark matter objects such as cosmic strings and domain walls.

\acknowledgments
We would like to thank the anonymous referee for noting an error in the signal and noise functions.
JJ is pleased to acknowledge the hospitality of the IPPP and their support via a DIVA fellowship.
SS and MS are supported by the UK Science and Technology Facilities Council (STFC) under grant ST/P001246/1.
During the revision of the manuscript, SS is funded by the Deutsche Forschungsgemeinschaft (DFG, German Research Foundation) -- 444759442.

\appendix

\section{The Detector Geometry of LISA and its Signal Response}
\label{app:geometry}

The shape and strength of a signal at LISA (or any other gravitational wave interferometer) induced by a gravitational perturbation due to a massive object in the vicinity of the detector, of course, depends on the distance, velocity and orientation of the latter with respect to the experiment.
In this appendix, we aim to define the relevant geometrical quantities that enter the derivation of the signal power spectrum associated to these events.

\begin{figure}[t]
\centering
	\includegraphics[width=0.65\columnwidth]{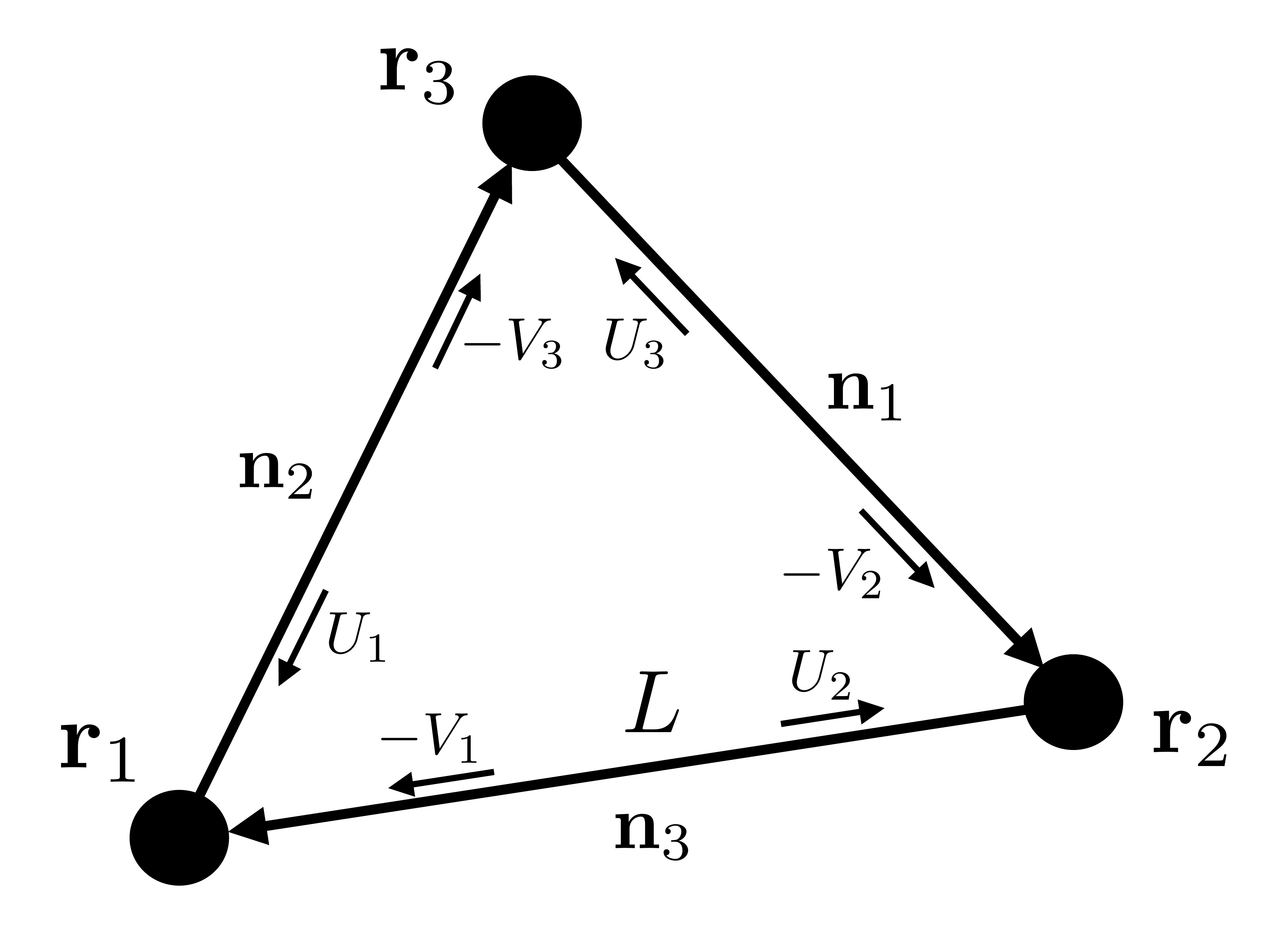}
	\caption{Schematic overview of the LISA interferometer. Here, $\vec{n}_i$ are unit vectors connecting pairs of satellites. The $U_i$ and $V_i$ encode the possible Doppler shifts of the laser beams that are exchanged between the detector nodes.}
\label{fig:LISAgeneral}
\end{figure}

In general, LISA is a space-based gravitational wave interferometer involving three distinct nodes, which are arranged in an equilateral triangle of side length $L \approx 2.5 \times 10^{6} \, \mathrm{km}$.
Following the notation of~\cite{Dhurandhar:2002zcl}, we schematically illustrate the general experimental setup of LISA in Fig.~\ref{fig:LISAgeneral}.
Each pair of nodes exchanges laser beams, such that, in principle, there are six functions, $U_{1,2,3}$ and $V_{1,2,3}$, that encode Doppler shifts due to gravitational perturbations by massive objects in the vicinity of the experiment.
The desired detector response to an acceleration burst is then given by a suitable linear combination of these functions.
Here, we use the so-called Michelson response function for the readout of a signal at a single detector node~\cite{Armstrong:1999},
\begin{equation}
\begin{split}
	X(t) =& U_1(t) + V_1(t) - U_1(t-2L/c) - V_1(t-2L/c) \\
	&- U_2(t-L/c) - V_3(t-L/c) \, ,
\end{split}
\label{eq:ResponseFunctionFullAppendix}
\end{equation}
where we have assumed that the arms of the interferometer are of equal length $L$.
The components of the response function are given by projections of the velocity perturbations onto the interferometer arms,
\begin{align}
	U_1(t) &= \vec{n}_2 \cdot \frac{\vec{v}_1(t)-\vec{v}_3 (t - L/c)}{c} \, ,\\
	V_1(t) &= \vec{n}_3 \cdot \frac{\vec{v}_1(t)-\vec{v}_2 (t - L/c)}{c} \, ,
\end{align}
and cyclic permutations thereof.
Here, the $\vec{n}_i$ denote the unit vectors pointing between two nodes, labelled by the opposite side of the triangle, $\vec{v}_i$ is the velocity perturbation of the $i$-th node induced by the gravitational pull and $c$ is the speed of light.
We note that other response function are also possible, see, e.g.,~\cite{Armstrong:1999,Dhurandhar:2002zcl}.

From the response function~\eqref{eq:ResponseFunctionFullAppendix} it is clear that any gravitational pull exerted by a massive object in the vicinity of the interferometer nodes has to be projected into the detector plane.
For example, in the extreme case, where only one satellite is accelerated perpendicular to the detector plane, e.g.~$\vec{n}_2 \cdot \vec{v}_1 = \vec{n}_3 \cdot \vec{v}_1 = 0$, the object will not leave any signature in the interferometer\footnote{Strictly speaking, this is not true, because the two other nodes will also experience a gravitational pull which is not perpendicular to the detector plane. Nevertheless, at large distances between the source and these nodes this effect is negligible.}.
Therefore, the detector response will depend on how an object traverses the detector volume, i.e.~on its orientation relative to the detector plane.

In the following, we want to define the relevant geometrical quantities describing this relative orientation of different macroscopic astrophysical objects we aim to probe with LISA.
For simplicity, we will only consider the close-approach limit, where the object passes by an interferometer node with an impact parameter smaller than the characteristic size of the detector, $D \lesssim L$.
As explained in the main text, in this case the gravitational perturbation of two of the three interferometer nodes can be neglected, such that the detector response function can be approximated by adding suitable time delays to~\cite{Vinet:2006fj}
\begin{equation}
    X(t) = -\vec{n}_i \cdot \frac{\vec{v}_i(t) - \vec{v}_i(t-4L/c)}{c} \, ,
\label{eq:ResponseApproximationAppendix}
\end{equation}
where $\vec{v}_i(t)$ is the velocity perturbation of the $i$-th node, that dominates compared to the two others.
In other words, $i$ is the interferometer node with smallest impact parameter with respect to the object traversing the detector volume.
Hence, we do not sum over the indices in~\eqref{eq:ResponseApproximationAppendix}.

\subsection{Spherical clumps}
\label{app:GeomSpheres}

\begin{figure}[t]
\centering
	\subfloat[Spherical clump and infinite string]{
		\centering
		\includegraphics[width=0.65\columnwidth]{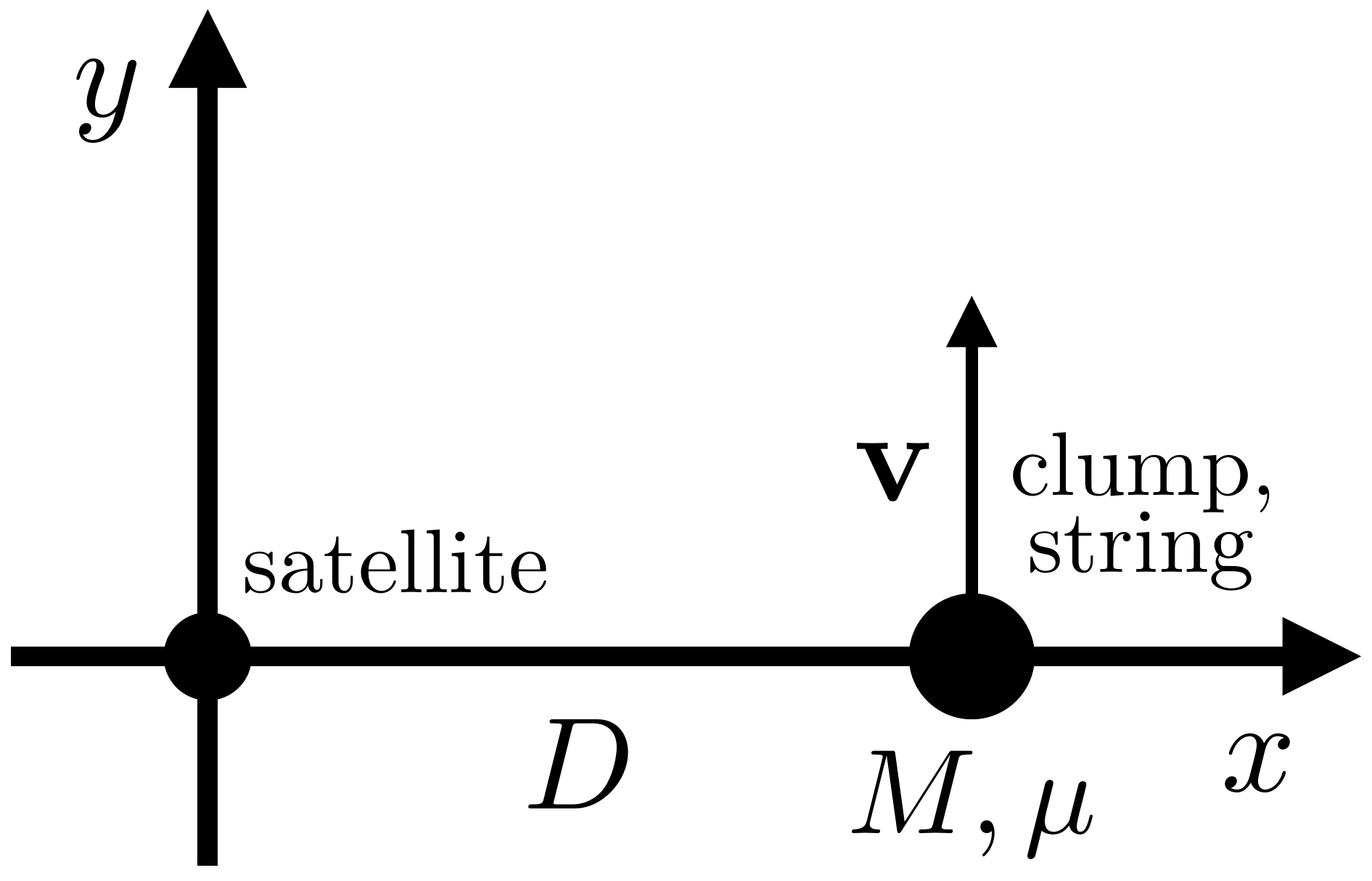}
	}
	\\
	\subfloat[Domain wall]{
		\centering
		\includegraphics[width=0.65\columnwidth]{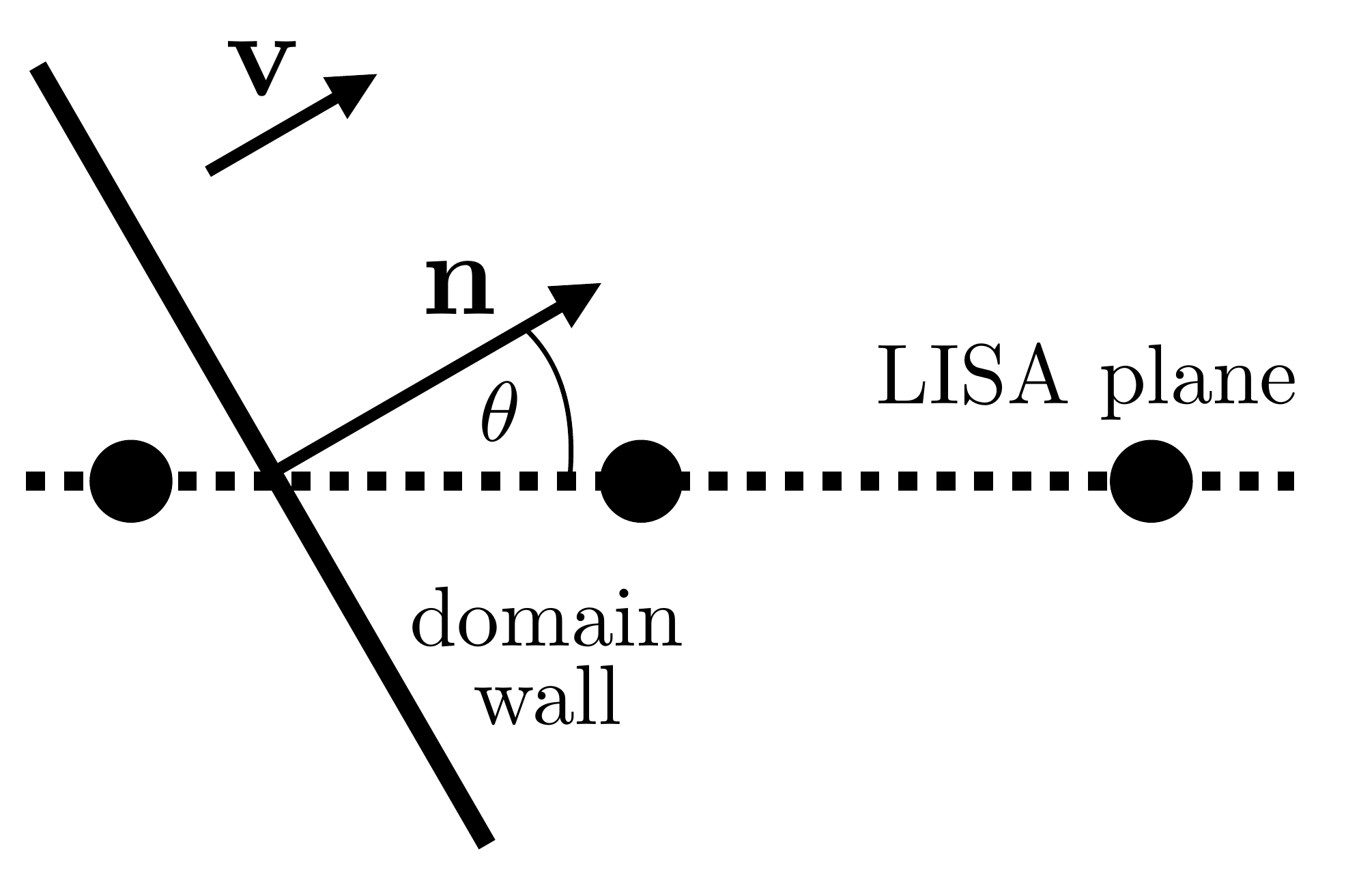}
	}
\caption{Schematic projections of a spherical clump and an infinite string (a) and a domain wall (b) passing by the nodes of the interferometer.
(a) The spherical clump of mass $M$ is chosen to move uniformly in parallel to the $y$-axis, i.e.~$\vec{v} = v \vec{e}_y$ with an impact parameter $D$. The infinite string of tension $\mu$ is parallel to the $z$-axis, such that it is perpendicular to the $xy$-plane. (b) The domain wall is travelling through the detector plane at an inclination angle $\theta$. Without loss of generality, we assume that the domain wall moves in the direction normal to it, $\vec{v}=v\vec{n}$.}
\label{fig:LISAobjects}
\end{figure}

Massive spherical objects, such as localized clumps of dark matter, are in a sense the most symmetric configuration when they pass by one of the LISA nodes.
That is, in the close-approach limit, their lack of an internal orientation allows to parametrize their motion relative to the detector by a velocity vector $\vec{v}$ and an impact parameter $D$, i.e.~the closest distance in an encounter between the massive clump and a detector node.
The former is completely determined by its magnitude $v$ and an arbitrary direction given in terms of two angles, i.e.~$\vec{v} = v \left( \sin\theta \cos\phi, \sin\theta \sin\phi, \cos\theta \right)$.
Clearly, not all of these four parameters will enter the detector response.
In fact, since the gravitational force between the spherical clump and the LISA satellite is only determined by their relative distance, we can consider the projection into the plane spanned by the satellite and the trajectory of the clump.
This effectively removes two degrees of freedom, such that we are left with the relative velocity $v$ and the impact parameter $D$.
As a particular example discussed in the main text, we can choose the spherical clump to be in a straight uniform motion with velocity $v$ parallel to the $y$-axis at an initial distance $D$ to the satellite.
The clump is furthermore confined to the $xy$-plane.
We illustrate this scenario in the top panel of Fig.~\ref{fig:LISAobjects}.

The above considerations determine the gravitational pull exerted by a massive clump on a single LISA satellite.
As pointed out earlier, in order to determine the corresponding detector response, this gravitational acceleration burst has to be projected into the detector plane.
The latter can be parametrized by an arbitrary unit normal vector, $\vec{n}_T = \left( \sin\vartheta \cos\varphi, \sin\vartheta \sin\varphi, \cos\vartheta \right)$.
In our analysis, the orientation of the detector plane is implemented in the detector response function, which in the close-approach regime, $D \ll L$, is given by~\eqref{eq:ResponseApproximationAppendix}.
That is, we can take this orientation into account by parametrizing the ``dominant" unit vector of the LISA triangle accordingly, $\vec{n}_i = \left( \sin\vartheta \cos\varphi, \sin\vartheta \sin\varphi, \cos\vartheta \right)$.
Note that, strictly speaking, we are slightly abusing notation here.
Obviously, the angles parametrizing the normal vector of the detector plane and the unit vector connecting two nodes of the triangle are not the same.
Nevertheless, since we will average over these angles later, we denote them by the same symbol to avoid an overload of notation.

In summary, in the close-approach limit, there are four geometrical degrees of freedom in total that enter the detector response to a localized massive clump travelling through the interferometer.
In particular, the clump's velocity $v$, the impact parameter $D$ as well as the orientation of the detector plane $(\vartheta, \varphi)$ completely determine the signal at LISA.
That is, the detector response is a function of these parameters, $X(t) = X(t, v, D, \vartheta, \varphi)$.

Finally, since we assume a locally isotropic situation, i.e.~the clumps can approach the interferometer from any direction equally likely, we uniformly average over the orientation of the detector plane in order to obtain the signal power spectrum from the detector response,
\begin{equation}
\begin{split}
    P(\omega) &= \avg{\abs{\tilde{X}(\omega, v, D)}^2} \\
    &= \frac{1}{4\pi} \int_{0}^{\pi} \diff \vartheta \sin\vartheta \int_{0}^{2\pi} \diff \varphi \abs{\tilde{X}(\omega, v, D, \vartheta, \varphi)}^2 \, .
  \end{split}
\label{eq:PowerSpectrumAngularAverage}
\end{equation}
Overall, the signal power spectrum then still depends on the velocity of the clump as well as the impact parameter of the encounter.

Let us close this discussion with a few words of caution.
Strictly speaking, the uniform average we have employed above, is not fully justified.
This is because the configuration we consider is not strictly isotropic.
Instead, there is a preferred direction in the system, given by the Sun, together with the detector, moving through the Universe.
In this sense, a uniform average is only an approximation.
We present a more detailed discussion of this in Appendix~\ref{app:dmvelocity}.

\subsection{Infinite strings}
\label{app:GeomStrings}

Unlike spherical clumps, or point-like particles, cosmic strings do have an internal degree of freedom.
That is, from a geometrical point of view, they are parametrized by line elements, thereby having an additional orientation themselves.
Obviously, when determining the detector response to a gravitational acceleration caused by a cosmic string, this orientation has to be taken into account.

In general, an infinite string can be parametrized by a straight line, $\vec{\gamma}(s) = \vec{x}_0 + s \vec{n}_\gamma$, where $\vec{n}_\gamma$ denotes an arbitrary unit vector.
In addition, the string can move in a direction with a certain velocity $\vec{v}$ relative to the nodes of the interferometer.
Again, similar to the case of spherical clumps, not all of these parameters will enter the detector response to a gravitational perturbation in the close-approach regime.
In fact, the gravitational field of the string only depends on the radial distance to the source, such that we can consider the projection into the plane perpendicular to the string.
As in the main text, we can choose a coordinate frame, in which the infinite string is parallel to the $z$-axis and the satellite located at the origin is, initially at $t_0=0$, at a minimum distance $D$ to the string.
We can then assume that the string is uniformly moving in a random direction in the $yz$-plane with velocity $v$, i.e.~$v_y = v \sin\theta$ and $v_z = v \cos\theta$, respectively.
Indeed, this reflects the fact that the string has an additional internal orientation as compared to spherical objects such as clumps.
This situation is depicted in the top panel of Fig.~\ref{fig:LISAobjects}.
Therefore, the gravitational acceleration burst induced by a cosmic string on a single LISA node depends on three parameters, namely the relative velocity $v$, the impact parameter $D$ as well as the direction of motion relative to the string orientation, parametrized by $\theta$.
Note that, equivalently, we could also choose a reference frame where the satellite is moving uniformly and the satellite is at rest.

Finally, as pointed out in the previous section, the overall gravitational acceleration has to be projected into the detector plane, parametrized by the angles $\vartheta$ and $\varphi$.
Therefore, in summary, LISA's detector response to a gravitational pull by cosmic string (in the close-approach regime) is a function of five geometrical parameters in total, $X(t) = X(t,v,D,\theta,\vartheta,\varphi)$.
In an isotropic Universe, the signal power spectrum is finally given by a uniform average over all arbitrary orientations involved,
\begin{equation}
\begin{split}
    P(\omega) =& \frac{1}{8\pi^2} \int_0^{2\pi} \diff\theta \int_0^\pi \diff\vartheta \sin\vartheta \\
    &\times \int_0^{2\pi} \diff\varphi \abs{\tilde{X}\left(\omega,v,D,\theta,\vartheta,\varphi\right)}^2 \, .
\end{split}
\end{equation}
In total, the signal power spectrum depends on the velocity of the string as well as the impact parameter of the encounter.
However, we note that, similar to the case of dark matter clumps, a uniform average over all possible directions might not be fully justified, see Appendix~\ref{app:dmvelocity}.

\subsection{Domain walls}
\label{app:GeomPlanes}

When determining the detector response of LISA to a gravitational potential sourced by an energy density localized on an infinite plane, i.e.~a domain wall, additional degrees of freedom compared to a spherical clump or an infinite string have to be taken into account.

Geometrically, the plane parametrizing a domain wall can be described by the algebraic equation $\vec{n} \cdot \left( \vec{r} - \vec{r}_0 \right) = 0$, where $\vec{r}_0$ is an arbitrary point in the plane and $\vec{n}$ denotes the unit vector normal to it.
Nevertheless, as we will see momentarily, the exact signal shape caused by a domain wall involves fewer geometrical parameters than, e.g., spherical objects or cosmic strings.
This is due to the fact that its gravitational field only induces a signal at LISA, if the domain wall is located in between the detector nodes, thereby separating them from each other\footnote{As discussed in the main text, this is because LISA only measures differential accelerations between the satellites. However, the gravitational field sourced by a domain wall does not depend on the distance, but is constant everywhere. Therefore, signals are only generated for configurations where the satellites are accelerated into opposite directions.}.
Therefore, we only have to consider a situation, where the triangle spanned by the LISA satellites intersects an infinite plane.
If the domain wall, or equivalently LISA, is moving at a certain velocity, this line of intersection will move, too, until it has completely passed the detector plane.
We illustrate this in the bottom panel of Fig.~\ref{fig:LISAobjects}.

The only geometrical quantities that enter the detector response function $X(t)$ in this scenario are, in fact, the orientation of the domain wall with respect to the triangle spanned by the LISA satellites as well as its relative velocity.
Without loss of generality, the former can be completely parametrized by, e.g., the unit vector normal to the plane, $\vec{n} = \left(\sin\theta \cos\phi, \sin\theta \sin\phi, \cos\theta\right)$, while we can assume the latter to point into the normal direction, $\vec{v} = v \vec{n}$, (cf.~Fig.~\ref{fig:LISAobjects}).
Accordingly, the detector response will be a function of these parameters only, $X(t) = X(t, v, \theta, \phi)$.
Finally, similar to the previous sections, in a locally isotropic dark matter distribution, the signal power spectrum associated to the gravitational perturbation by a domain wall traversing the detector volume is given by the uniform average over all possible orientations relative to the detector,
\begin{equation}
    P(\omega) = \frac{1}{4\pi} \int_{0}^{\pi} \diff \theta \sin\theta \int_{0}^{2\pi} \diff \phi \abs{\tilde{X}(\omega, v, \theta, \phi)}^2 \, .
\end{equation}
That means, the overall signal power spectrum will only depend on the velocity of the domain wall relative to the LISA detector.
However, we also note that, similar to the case of dark matter clumps, a uniform average over all possible directions might not be fully justified, as we will discuss in the following subsection.

\subsection{Velocity distribution of dark matter}
\label{app:dmvelocity}

In the previous subsections we have illustrated how our estimate of the signal power spectrum accounts for the relative orientation between the detector plane of LISA and the source of the gravitational perturbation.
In particular, we have assumed that the dark matter can approach the interferometer from any direction equally likely and hence uniformly averaged over the solid angle which parametrizes the latter (see, e.g.,~\eqref{eq:PowerSpectrumAngularAverage}).
Naively, this partly follows from the Maxwell-Boltzmann distributed velocities of the dark matter structures.
However, strictly speaking, this is not fully justified for the following reason.

In a naive approximation, it is reasonable to assume that the dark matter inside the halo surrounding our Galaxy behaves like an ideal gas of non-interacting particles and therefore roughly follows a Maxwell-Boltzmann distribution (see, e.g.,~\cite{Lisanti:2016jxe}),
\begin{equation}
    p(\vec{v}) = \left(\frac{1}{2 \pi v_0^2}\right)^{\frac{3}{2}} \exp \left( - \frac{\abs{\vec{v}}^2}{2 v_0^2} \right) \, ,
\label{eq:MaxwellBoltzmannVector}
\end{equation}
for some normalization $v_0$, which, from a microscopic perspective, is determined by the dark matter mass and the temperature of the gas.

The Maxwell-Boltzmann distribution~\eqref{eq:MaxwellBoltzmannVector} of the dark matter inside the halo of our Galaxy yields an isotropic uniform distribution for the direction in which the dark matter structures are moving.
That is, dark matter can approach the experiment from any direction equally likely.
In the previous sections, this feature is taken into account by a uniform average of the angles parametrizing the relative orientation between the detector plane of LISA and the trajectory of the dark matter structure (see, e.g.,~\eqref{eq:PowerSpectrumAngularAverage}).
Obviously, this is true in an isotropic reference frame where an observer is at rest inside the dark matter halo of the Galaxy.
However, in practice, the Sun, together with the detector, is moving through the halo at a constant velocity of $v_{\odot} \approx 220 \, \mathrm{km/s}$~\cite{GalacticConstants1986}, thereby imposing a \emph{preferred} direction on the system.
That means that not every direction in an encounter occurs equally likely, such that the average over these directions should not be uniform.
Instead, to correctly account for this one would need to weight the velocity in each direction according to the normal distribution~\eqref{eq:MaxwellBoltzmannVector} with the appropriate velocity shift by $v_{\odot}$, for example in $z$-direction, $v_z \to v_z - v_{\odot}$.

At first glance the situation looks even worse, because, in addition, LISA is moving on a complicated orbit around the Sun (see, e.g., Fig.~4 in~\cite{Danzmann2017LISA}).
However, this composite motion of the detector might turn out to be a blessing in disguise~\cite{Vinet:2006fj}, as it does not impose a single preferred direction but (at least naively) periodically changes the latter.
Hence, in order to account for the relative orientation between the experiment and the dark matter trajectory, a uniform average average over the orientation might indeed be closer to the experimental scenario than singling out only one preferred direction~\cite{Vinet:2006fj}.
In practice, as an approximation, we therefore take a uniform average over the solid angle accounting for the direction (see, e.g.,~\eqref{eq:PowerSpectrumAngularAverage}) and weight the detector response according to the probability distribution
\begin{equation}
    p(v) = \frac{1}{\sqrt{2\pi v_0^2}} \exp \left[ -\frac{1}{2} \left(\frac{v-v_{\odot}}{v_0}\right)^2\right] \, .
\end{equation}
Here, we try to approximate the motion of the detector through the dark matter halo with $v_{\odot} \approx 220 \, \mathrm{km/s}$~\cite{GalacticConstants1986} and finally normalize to the dark matter rms velocity of the latter, $v_0 = v_{\mathrm{rms}} / \sqrt{3}$ with $v_{\mathrm{rms}} \approx 270 \, \mathrm{km/s}$ (see, e.g.,~\cite{Kamionkowski:1997xg}).
Let us remark that, for the purpose of this work, we do not expect any large quantitative changes if a more accurate estimation of the dark matter velocity distribution with respect to the detector was performed.

\bibliography{references}

\end{document}